\documentclass[article, superscriptaddress, nofootinbib, showpacs,10pt]{revtex4}
\usepackage[titletoc,title,toc]{appendix}
\usepackage{graphicx}
\usepackage{amssymb}
\usepackage{epstopdf}
\usepackage{amsmath}
\usepackage{hyperref}
\usepackage{mathrsfs}
\usepackage{varioref}
\usepackage[active]{srcltx}
\expandafter\ifx\csname package@font\endcsname\relax\else
\expandafter\expandafter
\expandafter\usepackage
\expandafter\expandafter
\expandafter{\csname package@font\endcsname}%
\fi
\hypersetup{
    bookmarks=false,         
    pdfstartview={FitH},    
    colorlinks=true,       
    linkcolor=red,          
    citecolor=blue,        
    filecolor=blue,      
    urlcolor=blue           
}
\DeclareFontFamily{OT1}{pzc}{}
\DeclareFontShape{OT1}{pzc}{m}{it}%
             {<-> s * [0.900] pzcmi7t}{}
\DeclareMathAlphabet{\mathscr}{OT1}{pzc}%
                                 {m}{it}
\labelformat{section}{Section #1} 
\labelformat{subsection}{Section #1} 
\labelformat{equation}{Eq.(#1)} 
\labelformat{figure}{Fig.~#1} 
\labelformat{subfigure}{Fig.~\thefigure#1} 
\labelformat{table}{Tab.~#1} 
\newcommand{\be}{\begin{equation}}
\newcommand{\ee}{\end{equation}}
\newcommand{\bea}{\begin{eqnarray}}
\newcommand{\eea}{\end{eqnarray}}

\newcommand{\cosech}{\operatorname{cosech}}

\begin{document}

\title{Inertial non-vacuum states viewed from the Rindler frame}

\author{Kinjalk Lochan}%
\email{kinjalk@iucaa.ernet.in}%
\author{T. Padmanabhan}%
\email{paddy@iucaa.ernet.in}
\affiliation{IUCAA, Post Bag 4, Ganeshkhind,\\
Pune University Campus, Pune 411 007, India}

\begin{abstract}

The appearance of the inertial vacuum state in Rindler frame has been extensively studied in the literature, both from the point of view of quantum field theory developed using Rindler foliation and using the response of an Unruh-Dewitt detector. In comparison, less attention has been devoted to the study of inertial \textit{non-vacuum} states when viewed from the Rindler frame. We provide a comprehensive study of this issue in this paper. We first present a general formalism describing the characterization of arbitrary inertial state (i) when described using an arbitrary foliation and (ii) using the response of an Unruh-DeWitt detector moving along an arbitrary trajectory. This allows us to calculate the mean number of particles in an  arbitrary inertial state, when the QFT is described using an arbitrary foliation of spacetime or when the state is probed by a detector moving along an arbitrary trajectory. We use this formalism to explicitly compute the results for the Rindler frame and uniformly accelerated detectors.

 Any arbitrary inertial state will always have a thermal component in the Rindler frame with additional contributions arising from the non-vacuum nature. We classify the nature of the additional contributions in terms of functions characterizing the inertial state. We establish that for all physically well behaved normalizable inertial states, the correction terms decrease rapidly with the energy of the Rindler mode so that the high frequency limit is dominated by the thermal noise in any normalizable inertial state. However, inertial states which are not strictly normalizable like, for example, the one-particle state with definite momentum, lead to a constant contribution at all high frequencies in  the  Rindler frame. We show that a similar behavior arises in the response of the Unruh-DeWitt detector as well. In the case of the detector response, we provide a physical interpretation for the constant contribution at high frequencies in terms of total detection rate of co-moving inertial detectors. We also describe two different approaches for defining a  transition \textit{rate} for the Unruh-DeWitt detector, when the two-point function lacks the time translation invariance and discuss several features of different definitions of transition rates. The implications are discussed.

\end{abstract} 

\maketitle

\section{Introduction}

Quantum field theory (QFT) in curved spacetime (or in non-inertial co-ordinates) has traditionally introduced new conceptual features into the standard notions of QFT, developed in the inertial co-ordinates in flat spacetime. One of the key results is that neither the particle content of the quantum states, nor the structure of vacuum fluctuations remain generally covariant \cite{CURVEDQFT}. Given a field in the spacetime, the particle content of the field can be evaluated {\it only after} introducing  some  mode functions into which  the field can be decomposed. Such a mode decomposition is usually selected by the  ``positive frequency" criterion in a particular

 (1+3) foliation of the spacetime. Generically, different congruences of observers can be associated with differnet (1+3) foliations of spacetime by, say, identifying the normal vector to the spacelike surfaces with the four-velocity  of the congruence of observers. Every congruence of observers will interpret the particle content of the field  in terms of modes which appear natural to her/him. Consequently, we are led to different sets of mode functions and creation/annihilation operators, related between different sets, in general, by Bogoliubov transformations. A vacuum state defined by one set of annihilation operators will be ---in general--- perceived as a non-vacuum state  when interpreted using a different set of mode functions. A well-known example is the vacuum state in Minkowski foliation which appears as a thermal bath to  uniformly accelerating observers. Similarly  the vacuum state of a geodesic observer close to the horizon of the black holes will appear as a thermal state to an asymptotic stationary observer, who is an accelerating observer with respect to the in-falling geodesic observer.

Another ---and possibly more direct--- probe of the quantum field (closely related to mode functions and foliations but not identical to it) is coded in its interaction with particle detectors. A detector moving along a specific trajectory and

coupled to the field will respond in a manner specific to the trajectory. Studying the response of such detectors gives another characterization of the quantum field. It is usual  to study the response of detectors linearly coupled to the quantum field, viz. the Unruh-DeWitt detector \cite{Unruh}. Depending upon different trajectories it takes, the detector will click at different rates. The detector response depends essentially on the two-point (Wightman) function of the field evaluated along the trajectory. That is, while the particle content depends on the mode decomposition, the detector response is related to the  two-point function along its trajectory and ---a priori--- there is no reason to expect the detector response to match with the particle content  of the field defined using foliation dependent mode functions. In fact, they will \textit{not} match with each other in a generic case  (see e.g. \cite{DetvsBGLB}). However, in certain cases, (e.g. detector on a Rindler trajectory with the field being in the Minkowski vacuum) the number of ``clicks'' the detector registers  is directly related to  the number expectation value obtained by the mode functions natural to the foliation. One important property of the inertial vacuum state is that the two point function, for all those observers who move along trajectories which are integral curves of time-like Killing fields, is time-translation invariant. In those cases, the Unruh-DeWitt detector clicks at a \textit{uniform} rate and the particle content of the field is proportional to the detector response.

In this paper, we extend the above analysis to the situation in which the quantum field is in a \textit{non-vacuum} state. We develop a formalism to study both the particle content and the detector response for the field in terms of a single entity defined as an {\it effective field}. We first show that all the  information  relevant to a particular foliation (used by congruences of observers) or a detector in a particular trajectory can be extracted from this effective field. 

Within this set-up, we investigate the cases in which the two point function (generically) lacks the property of time translation invariance. For example, this happens when the state of the field is not the inertial vacuum or, more specifically, when the inertial state is not a  momentum eigenstate for a particular mode. We first develop a general set up for evaluating interesting quantities like number expectation value, detector transition probability, detection rates etc. in the  general context. Moreover, since the behavior of the vacuum state  is already well-known in the literature we focus our attention on the departures from the vacuum response, when the field happens to be in a non-vacuum state.  We study the quantum filed when it is not in vacuum state in both the approaches, i.e. the particle content and the detector response. In a non-vacuum state a Unruh-DeWitt detector will ``click'' also in the inertial frame, thereby making the identification of the genuine effects of non-inertial frames somewhat non-trivial; we take special care to isolate this feature whenever applicable.

As a concrete demonstration of these ideas, we show that an  arbitrary inertial quantum  state of Minkowski mode, when studied from the Rindler frame,  will give rise to a thermal part plus additional corrections. We concentrate on the additional  corrections in detail using our formalism. We establish that for any arbitrary state in the Minkowskian frame, the corrections, decay down rapidly with the energy scale the Rindler observer uses to probe the field. In other words, for any arbitrary state in the inertial frame, the state of the field looks largely thermal when the observers probe it at scales larger than the one  characterizing the Rindler frame, viz. $a$, i.e.,  $\Omega \gg a$. Therefore, the thermal behavior as seen by Rindler observers persists for sufficiently energetic probes.

We also characterize the decay profile of the corrections, in terms of a suitably defined Minkowski distribution function, which characterizes the particle content of the state in the Minkowski frame. We show that a red-shifted version of the Minkowski distribution function determines the corrections to the thermal spectrum. A similar trend is shown by the Rindler detector as well. Interestingly, although  the two point function is not time translation invariant for the Rindler observer, the detector and the number expectation value show similar type of corrections (if the detector is kept ``on'' for sufficiently large duration).

This general formalism will be particularly useful while dealing with the  cases such as pair production from electric field, particle production in an expanding  universe and so on, when the initial state under evolution does not happen to be the vacuum state. We will be dealing with those cases in a subsequent work.  Another  important motivation for for such a study is the possible extension for the black hole when the collapse of matter to form a black hole occurs  with the  field being in  a non-vacuum state. Insights gained here from Rindler-Minkowski dictionary can  be applied to study black hole evaporation in a non-vacuum state in a more or less in a straightforward manner.  We will discuss the spectral distortion of the black-hole radiation, elsewhere.

The  study in this paper might also have some relevance for the emergent gravity paradigm \cite{EmergentGravity}. In this approach gravity, as we understand it, is attributed with a thermodynamic status using the thermodynamic nature of Killing horizons since one can envisage horizons as observed by local Rindler observers at each event. Therefore, characterizing the non-vacuum state of the quantum field from the perspective of the Rindler observer holds some importance in this case as well.

We organize the manuscript in the following manner. First in section II, we set up the formalism for characterizing the correction (with respect to the vacuum state) in the number expectation values  associated with the modes corresponding to an arbitrary foliation. We show that all the relevant information about the state and the trajectory is captured in a hybrid effective field, using which we can calculate the number expectation value for any trajectory. Moreover, we can classify the detector response using this effective field.  In section III, we use this formalism to classify the number expectation value of arbitrary Minkowski states when viewed by the Rindler observer.  We show that the correction over the vacuum part decays rapidly with increasing energy of the probe. The rate of decay is closely related to the distribution in terms of Minkowski modes. Section IV deals with the Rindler detector response for non-vacuum Minkowski state, where we discuss some salient features of the transition rates. In section V, we discuss two different approaches for defining the transition rate and note that the infinite time detector transition rate for one of the approaches can be viewed as a Wigner function.  Similarly the finite time detector will also show a transition rate which, for all finite energy states of Minkowski frame, remains finite. 

We summarize our main points  and discuss their possible extension in section VI.

\section{Formalism}

Throughout the paper we consider a massless, real, scalar field with minimal coupling. (However, the scheme and the results developed here are general enough to include other kinds of massive fields in a straight forward manner.) We will keep the dimensions of spacetime unspecified in general, but will go to the (1+1) dimensional case for explicit demonstrations for technical convenience. Similar results are obtainable in (1+3) dimensions as well.

The equation of motion for a massless real scalar field minimally coupled to gravity, will be the Klein Gordon (KG) equation

\bea
\square \phi =0, \label{KG}
\eea

where the operator $\square$ is written using covariant derivative operator $\triangledown_{\mu}$ compatible with the spacetime metric $g_{\mu \nu}$.

There exists a covariant notion of inner product between solutions of the KG equation, defined by: 

\bea
(f,h)\equiv i \int d^{d} x |g|^{1/2} g^{0 \nu} f^*({\bf x},t)\overleftrightarrow{\partial}_{\nu} h({\bf x},t). \label{innerProduct}
\eea

If the functions  are well-behaved asymptotically, this inner product remains time invariant.

Furthermore,

\bea
(f,h)^*=(h,f)=-(f^*,h^*). \label{innerProductRules}
\eea

We next introduce a complete set of 

orthonormal modes satisfying \ref{KG}, and:

\bea
(u_k,u_{k'})=\delta_{kk'}; \hspace{0.5 in} (u_k,u^*_{k'})=0; \hspace{0.5in}(u^*_k,u^*_{k'})=-\delta_{kk'}.
\eea

Any solution to 

\ref{KG} can be expanded as a linear combination of such modes.

Therefore, the field operator can be written in terms of the mode functions $u_k(x)$ as

\bea
\hat{\phi}(x)=\sum_k\left(\hat{a}_k u_k + \hat{a}_k^{\dagger} u_k^*\right).
\eea

with the usual creation/annihilation operators, associated with this particular mode functions. The functions $u_k$  can also be thought of as the mode functions for a particular congruence of of observers (not necessarily Minkowskian) associated with a specific foliation of the spacetime. 

An arbitrary single excitation state corresponding to modes $u_k$, is defined by:

\bea
|\Psi \rangle =\sum_k {\cal F}_k \hat{a}_k^{\dagger} |0\rangle_u, \label{1PSDiscrete}
\eea

where ${\cal F}_k$ denotes the probability amplitude of the distribution, constrained by the normalization condition 
$\sum_k|{\cal F}_k|^2 =1$.
For convenience we have considered this single-excitation state\footnote{ We consider a well defined state which is partially localized in both position as well as the momentum space. The particle states in the Fock basis will be the cases where ${\cal F}_k \propto \delta_{k k'}$, which, of course, will be non-normalizable.}, however,  similar analysis can be done for a general multiply excited state, an example of which we will discuss in a later section. In fact a multiply excited state can be viewed as a direct product state over various single excitation stateswhich allows a direct generalization.

The two point correlation function of the  field in this state is given by:

\bea
\bar{C}(x,y)\equiv \langle \Psi| \hat{\phi}(x)\hat{\phi}(y) |\Psi \rangle ={}_u\langle 0| \hat{\phi}(x)\hat{\phi}(y) |0 \rangle_u + \Phi_{\text{eff}}(x) \Phi^*_{\text{eff}}(y)+\Phi^*_{\text{eff}}(x) \Phi_{\text{eff}}(y), \label{effectiveCF0}
\eea

where we have defined an \textit{effective field} corresponding this state by:

\bea
\Phi_{\text{eff}}(x)=\sum_k {\cal F}_k u_k (x) \label{effectivefield}.
\eea

We see that $\bar{C}(x,y)$ has a vacuum part and a correction term. Since, we will be more interested in the corrections to the vacuum contribution, when the inertial state does not happen to be the vacuum corresponding to $u_k$ modes, it is convenient to work with 

\bea 
{\cal C}(x,y) \equiv \bar{C}(x,y) -{}_u\langle 0| \hat{\phi}(x)\hat{\phi}(y) |0 \rangle_u=\Phi_{\text{eff}}(x) \Phi^*_{\text{eff}}(y)+\Phi^*_{\text{eff}}(x) \Phi_{\text{eff}}(y), \label{effectiveCF}
\eea

which may be called the {\it effective correlation function}.
From the effective correlation function we can obtain the number expectation value in a particular mode
by a simple relation: 

\bea
n_{\ell}=\langle \hat{a}_{\ell}^{\dagger}\hat{a}_{\ell} \rangle =-(u^*_{\ell}(x),(u_{\ell}(y), {\cal C}(x,y))), \label{NEXFormal}
\eea
where $\langle \hat{O} \rangle$ denotes expectation value of the operator $\hat{O} $ in an arbitrary state $|\Psi \rangle$.

Using the standard properties of the correlation functions and the modes (viz. their satisfying \ref{KG} and vanishing sufficiently fast at large distances so that the integrals are well-defined) we can show that the expression \ref{NEXFormal} is time independent and counts the number of particles in mode $l$.

We can rewrite the above expression in terms of the effective field \ref{effectivefield} as

\bea
\langle \hat{a}_{\ell}^{\dagger}\hat{a}_{\ell} \rangle = |{\cal F}_{\ell}|^2 = |(u_{\ell}(y), \Phi_{\text{eff}}(y))|^2. \label{Nex001}
\eea

Let us now consider what happens when we decide to introduce another set of mode functions $v_k$

which could be thought of as being associated with another foliation and another congruence of observers. These  modes $v_k$ will be  related to the original modes 
$u_k$ via Bogoliubov transformation:

\bea
v_j=\sum_i(\alpha_{ji}u_i+\beta_{ji}u^*_i). \label{BogoliubovCoeff}
\eea

The field operator, expressed in terms of these new modes, will be:

\bea
\hat{\phi}(x)=\sum_k\left(\hat{A}_k v_k + \hat{A}_k^{\dagger} v_k^*\right),
\eea

where $\hat{A}_k$ is the annihilation operator corresponding to mode $v_k$, which also can be written in terms of the annihilation and creation operators of the modes $u_k$ as

\bea
\hat{A}_k = \sum_i (\alpha^*_{ki}\hat{a}_i-\beta^*_{ki}\hat{a}_i^{\dagger}). \label{BogoliubovOpertor}
\eea

Clearly the number eigenstates of $\hat{a}_k^{\dagger}\hat{a}_k$ are not the eigenstates of the operator $ \hat{A}_k^{\dagger}\hat{A}_k$. One of the key quantities we are interested in is the expectation value of this quantity in our quantum state. We will now find several useful expressions for the same.

(i) To begin with, the number expectation value in a mode ${\ell}$ (as defined using in $v_k$s) can be written in terms of the correlation function as:

\bea
\langle \hat{A}_{\ell}^{\dagger}\hat{A}_{\ell} \rangle =  -(v^*_{\ell}(x),(v_{\ell}(y), {\cal C}(x,y))). \label{NExpectation}
\eea

The above expression relates the number expectation value in a particular mode, associated with, say, the choice of  a specific foliation  to the two point correlation function in any arbitrary state of the field. 

In this approach we think of  ${\cal C}(x,y)$ as fundamental and encoding information about the original state. Obviously, ${\cal C}(x,y)$ as well as the dot products in this equation are covariant; so the foliation dependence arises only through the choice of the mode functions.

(ii) Using \ref{effectiveCF} we can again rewrite the above expression as

\bea
\langle \hat{A}_{\ell}^{\dagger}\hat{A}_{\ell} \rangle =  -[(v_{\ell}(y), \Phi^*_{\text{eff}}(y))(v^*_{\ell}(x),\Phi_{\text{eff}}(x))+(v_{\ell}(y), \Phi_{\text{eff}}(y))(v^*_{\ell}(x),\Phi^*_{\text{eff}}(x))], \label{generalNEX}
\eea

Using the relations \ref{innerProductRules} and taking into account that the spatial variables are summed over in \ref{innerProduct}, i.e. act as dummy variables,

we can write \ref{generalNEX} as

\bea
\langle \hat{A}_{\ell}^{\dagger}\hat{A}_{\ell} \rangle = [\left|(v_{\ell}, \Phi^*_{\text{eff}})\right|^2 + \left|(v_{\ell}, \Phi_{\text{eff}})\right|^2]. \label{mostgeneralNEX}
\eea

In this approach, the effective field  $\Phi_{\text{eff}}$ encodes the information about the state of the field. In a way, this is a more fundamental description since ${\cal C}(x,y)$ can be expressed in terms of $\Phi_{\text{eff}}$  as shown in \ref{effectiveCF}, but not the other way around, as we shall soon see.

 Clearly, the expression \ref{mostgeneralNEX} reduces to \ref{Nex001} when the modes $v\equiv u$.

(iii) In terms of the Bogoliubov coefficients \ref{BogoliubovCoeff} we can express the correction to the number expectation value as

\bea
\langle \hat{A}_{\ell}^{\dagger}\hat{A}_{\ell} \rangle =\left [\left|\sum_i \beta_{{\ell}i} {\cal F}_i  \right|^2 +\left|\sum_i \alpha^*_{{\ell}i} {\cal F}_i  \right|^2\right]. \label{NEXinBogoulibov}
\eea

\noindent As we shall see from the ensuing discussion, the effective field $\Phi_{\text{eff}}$  will turn out to be a very useful entity. 

Given the relations

\ref{innerProductRules} we can also show that 

\bea
(\Phi_{\text{eff}},\Phi_{\text{eff}})=1; \quad (\Phi_{\text{eff}},\Phi^*_{\text{eff}})=0; \quad 
(\Phi^*_{\text{eff}},\Phi^*_{\text{eff}})=-1.
\eea

for physically normalizable states.

Therefore,

\bea
(\Phi_{\text{eff}}(x), {\cal C}(x,y)) &=& \Phi^*_{\text{eff}}(y), \nonumber\\ 
(v_{\ell}(x), {\cal C}(x,y)) &=& \left(\sum_i \alpha^*_{{\ell}i} {\cal F}_i\right)\Phi^*_{\text{eff}}(y)- \left(\sum_i \beta^*_{{\ell}i} {\cal F}^*_i\right)\Phi_{\text{eff}}(y).
\eea

In terms of this auxiliary field and the modes corresponding to this field, we can in fact express the quantum state itself as

\bea
|\Psi\rangle = \sum_k (u_k, \Phi_{\text{eff}}) \hat{a}_k^{\dagger} |0\rangle_u.
\eea

Thus, the correction over the vacuum contribution is completely characterized by an effective field $\Phi_{\text{eff}}$ from which we can reconstruct the two-point function and the state uniquely. However, given a two point function we can not uniquely fix the state. Thus, the field $\Phi_{\text{eff}}$ contains some additional  information. For $n-$ tuple excitation state, $n$ such fields will be required to describe the state uniquely (see Appendix A).

We will next discuss the response of the detectors in light of this kind of effective correlation function. This will help us to compare the results based on the foliation  with the detector response. A generalization of these ideas for an arbitrary state is presented in Appendix A.

\subsection{Detector Response}

In quantum field theory, an operational definition of a particle is `something which a detector coupled with the field detects'.

Consider an  Unruh-deWitt detector,  coupled linearly to a scalar field $\phi(x)$, that moves on a world-line $x^{\mu}(\tau)$ (where $\tau$ is the proper-time along the trajectory) and is described by the interaction Lagrangian

\bea
{\cal L}_{\text{int}} = c \hat{m}(\tau) \phi[x(\tau)], \label{InteractionLagrangian}
\eea

where $c$ stands for the linear coupling constant, $ \hat{m}(\tau)$ is the time-dependent detector monopole operator.  The probability of transition from

state $|\Psi_u, E_0\rangle $ ( field is in the state $|\Psi\rangle_u$, the detector is in its ground state) to any state in which the detector clicks

is given, in first order perturbation theory, by: 

\bea
\bar{P} =|c|^2\int_{-\infty}^{\infty}\int_{-\infty}^{\infty} d\tau d\tau'\sum_E e^{i(E-E_0)(\tau-\tau')}|\langle E|\hat{m}(0)|E_0\rangle |^2 
{}_u\langle\Psi |\phi[x(\tau')]\phi[x(\tau)]|\Psi\rangle_u. \label{ProbVac}
\eea

Thus, the detector response depends crucially on the two point function of the field for the state.  We will be interested in the contributions over and above the vacuum contribution (e.g. the thermal part for Rindler observer \cite{CURVEDQFT}) and hence we will drop the contribution from vacuum state in our analysis. Also, for additional clarity we only consider a two-state detector. For given set of modes $u_k$, we can define the effective field $\Phi_{\text{eff}}$ as in \ref{effectivefield} using which the 

the transition probability can be written as:

\bea
P =|c|^2\int_{-\infty}^{\infty}\int_{-\infty}^{\infty} d\tau d\tau' e^{i(E-E_0)(\tau-\tau')}|\langle E|\hat{m}(0)|E_0\rangle |^2 \left[\Phi_{\text{eff}}(x[\tau]) \Phi^*_{\text{eff}}(y[\tau'])+\Phi^*_{\text{eff}}(x[\tau]) \Phi_{\text{eff}}(y[\tau'])\right],
\eea

which in fact is a sum of squared amplitude of Fourier Transforms

\bea
P =|c|^2 |\langle E|\hat{m}(0)|E_0\rangle |^2\left[\left|\int_{-\infty}^{\infty} d\tau \Phi_{\text{eff}}(\tau)e^{i(E-E_0)\tau}\right|^2+\left|\int_{-\infty}^{\infty} d\tau \Phi^*_{\text{eff}}(\tau)e^{i(E-E_0)\tau}\right|^2\right],
\eea

of the effective field (and its complex conjugate) when viewed in terms of observer's time. (We have subtracted out the contribution from the vacuum state; we will not mention this fact specifically in the further analysis).

A physically more realistic system will have the interaction \ref{InteractionLagrangian} switched on only for a finite duration on the trajectory of the 

detector, say from $\tau=T_0$ till $\tau=T$. In that case, the transition probability expression will be modified to

\bea
P =|c|^2 |\langle E|\hat{m}(0)|E_0\rangle |^2\left[\left|\int_{T_0}^{T} d\tau \Phi_{\text{eff}}(\tau)e^{i(E-E_0)\tau}\right|^2+\left|\int_{T_0}^{T} d\tau \Phi^*_{\text{eff}}(\tau)e^{i(E-E_0)\tau}\right|^2\right], \label{1stOrderTransition}
\eea

apart from a finite duration vacuum contribution. 

Using the new quantity $\Phi_{\text{eff}}(\nu;T_0,T)$ defined by \footnote{Note that in $T_0\rightarrow -\infty$ and $T\rightarrow\infty$ limit, $\Phi_{\text{eff}}(\nu;T_0,T)$ is just the inverse Fourier transform of $\Phi_{\text{eff}}(\tau)$ w.r.t $\nu$, which we denote as ${\cal F}_{\nu}$.}

\bea
\Phi_{\text{eff}}(\nu;T_0,T) =\int_{T_0}^T d\nu \Phi_{\text{eff}}(\tau)e^{i\nu \tau}, \label{TdepFT}
\eea

the expression \ref{1stOrderTransition} can be written as

\bea
P =|c|^2 |\langle E|\hat{m}(0)|E_0\rangle |^2\left[\left| \Phi_{\text{eff}}((E-E_0);T_0,T)\right|^2+\left| \Phi_{\text{eff}}(-(E-E_0);T_0,T)\right|^2\right].\label{1stOrderTransition2}
\eea

It is useful to compare the above expressions with the standard result in time dependent perturbation theory of nonrelativistic quantum mechanics. In that case,

 the probability of transition from a state $m$ to a state $n$ (eigenstates of unperturbed Hamiltonian) by the action of a time dependent potential $ \hat{V}(t)$ is given by the text book expression:

\bea
|c |^2 = \int_{T_0}^{T}\int_{T_0}^{T}dt dt' e^{i\omega_{nm}(t-t')}\langle n|\hat{V}(t)|m \rangle \langle m|\hat{V}(t')|n \rangle = \left|\int_{T_0}^{T} dte^{i\omega_{nm}t}\langle n|\hat{V}(t)|m \rangle\right|^2,
\eea

which, in the spirit of \ref{TdepFT} can be rewritten as

\bea
|c |^2 = \left|\langle n|\hat{V}(\omega_{nm};T_0,T)|m \rangle\right|^2.
\eea

Therefore, structurally the expression \ref{1stOrderTransition} or  \ref{1stOrderTransition2} is similar to a case of a non-relativistic system with our effective field $\hat{m}\Phi_{\text{eff}}(t)$ playing the role of  a time dependent external potential. In the relativistic theory, its   complex conjugate also makes contribution which can be ultimately traced to the fact that Schr\"{o}dinger eigenfunctions have only positive frequency evolution $\exp(-iEt)$ while the KG modes have both positive and negative frequency parts.

From this expression we can also calculate the total transition \textit{rate} as the derivative of \ref{1stOrderTransition} with respect to $T$

\bea
R =|c|^2 |\langle E|\hat{m}(0)|E_0\rangle |^2\left[\Phi_{\text{eff}}(T)e^{i \Delta E T} \int_{T_0}^{T} d\tau \Phi^*_{\text{eff}}(\tau)e^{-i\Delta E \tau} +\Phi^*_{\text{eff}}(T)e^{i\Delta E T}\int_{T_0}^{T} d\tau \Phi_{\text{eff}}(\tau)e^{-i\Delta E \tau}+\text{c.c.}
\right],\nonumber\\ \label{RateFormal}
\eea

where $\Delta E =E-E_0$. The exact $\tau$ dependence of $\Phi_{\text{eff}}$ is known once the state and the modes corresponding to the observer are specified.

The  excitation rate of the detector can then be used to obtain an expression for the mean number of particles present in the state \cite{CURVEDQFT}. As to be expected, this will --- in general ---  be different from the result obtained from \ref{NEXinBogoulibov}. However, it is possible that in some special cases these two quantities match, even though the  correlation function is  not invariant under time translation. In such cases, we can define a time dependent transition rate.

We will now consider a specific demonstration of these corrective terms for (i) the number expectation value, (ii) the transition probability and (iii) the transition rate, in the case of Minkowski and Rindler observers. These expressions are extensively discussed in the literature  for the Minkowski \textit{vacuum state}. But the corresponding results for the  \textit{non-vaccum} states have attracted very little attention. (We could find only \cite{AudretschMuller} and \cite{PaddyTP} for instance, which are somewhat similar in spirit to the current discussion and constitute special cases of the general results presented here.) This will be the main thrust of the discussion in the forthcoming sections. We will concentrate on singly excited states; the generalization to $n$-tuple excitation (as well as the most general superposed state) is straightforward and is  presented in Appendices.

\section{Inertial excited states observed in Rindler frame}

\subsection{Single excitation state}

Since we define the state in terms of a Minkowski foliation, we want the state to be well behaved with respect to the symmetries in this context.

Therefore, we will work with normalized Lorentz invariant states, so that a single excitation state in $(1+3)$ dimensions can  be written as:

\bea 
\vert \Psi \rangle \equiv \vert 1 \rangle =\int \frac{d^3  {\bf k}}{(2\pi)^{3/2}} \frac{1}{\sqrt{2\omega_{| {\bf k}|}}}f({\bf k}) \hat{a}^{\dagger}( {\bf k})|0\rangle, \label{1PS}
\eea

where $f({\bf k})$ satisfies the normalization condition

\bea
\int \frac{d^3  {\bf k}}{(2\pi)^{3}} \frac{1}{2\omega_{| {\bf k}|}}|f({\bf k})|^2 =1.
\eea

This state $\vert 1 \rangle$ is clearly a superposition of one particle states in the Fock basis with the amplitude for the state to be found to have a particle of momentum $\bf k$ being specified by the function $f( {\bf k})$ which completely specifies the state.

When the modes are labeled by a continuous index, we can use \ref{1PSDiscrete} with the identification: 

\bea
{\cal F}_k \rightarrow \frac{f({\bf k})}{\sqrt{2\omega_{| {\bf k}|}}}.
\eea

The number expectation value for a particular mode (labeled by the vector, say ${\bf j}$), evaluated in this superposed state, is given by

\bea
\langle 1 \vert  \hat{a}_{\bf j}^{\dagger}\hat{a}_{\bf j}\vert 1 \rangle= \frac{|f({\bf j})|^2}{(2\pi)^{3}2\omega_j}.
\eea

Further, this state is an eigenstate of the total number operator (which counts the total number of particles across different modes)

\bea
\hat{N} \vert 1 \rangle = \int  d^3  {\bf k} \hspace{0.05 in} \hat{a}_{\bf k}^{\dagger}\hat{a}_{\bf k} \vert 1 \rangle = \vert 1\rangle.
\eea

The effective field $\Phi_{\text{eff}}$ for the Minkowskian modes is given as

\bea
\Phi_{\text{eff}}(x)=\int \frac{d^3  {\bf k}}{(2\pi)^{3}} \frac{1}{2\omega_{| {\bf k}|}}f( {\bf k}) e^{i k\cdot x}. \label{eff4mink}
\eea

For technical convenience we will now switch to $(1+1)$ dimensions, with signature $(-,+)$. Writing in terms of frequency modes in $(1+1)$ dimension ($t, x$), we get

\bea 
\vert 1 \rangle = \int_0^{\infty} \frac{d \omega}{\sqrt{4 \pi \omega}}f(\omega)\hat{a}^{\dagger}(\omega)|0\rangle, \label{1PS2}
\eea

where 

\bea
f(\omega)\hat{a}^{\dagger}(\omega)=f(k_x) \hat{a}^{\dagger}(k_x)+f(-k_x) \hat{a}^{\dagger}(-k_x).
\eea

Thus the Minkowski modes correspond to the $u$ modes considered in section II and the state will be first specified in the Minkowski modes and then will be studied in terms of Rindler modes which serve as the $v$-modes. For $(1+1)$ dimensional calculations we later on drop the subscript $x$ from $k_x$ while realizing that $k$ is a continuous variable taking value along the real line.

We first work with the state \ref{1PS} or \ref{1PS2} but later go on to see that our conclusions hold for a generic multiple excitation states and the corresponding superpositions thereof (Appendix B, C).

The number expectation value for the Rindler observer at frequency $\Omega$ in this state is given as

\bea 
 \bar{N}_{\Omega} = \langle \Psi| \hat{A}^{\dagger}(\Omega)\hat{A}(\Omega)|\Psi\rangle.
\eea

It is fairly easy exercise to show that the number expectation value always has a thermal component just as it would have been for the vacuum case. As usual, we will  subtract that part out and concentrate only upon the distortions brought to the thermal spectrum by the non-vacuum state.

Using continuum version of the expression \ref{NEXinBogoulibov}, written in terms of Rindler modes, the correction term can be expressed as

\bea
N_{\Omega}
=
\left[
\left|\int_0^{\infty} \frac{d \tilde{\omega}'}{\sqrt{4\pi\tilde{\omega}'}}\alpha^*_{\Omega \tilde{\omega}'}f(\tilde{\omega}')\right|^2+\left|\int_0^{\infty} \frac{d \tilde{\omega}}{\sqrt{4\pi\tilde{\omega}}}\beta_{\Omega \tilde{\omega}}f(\tilde{\omega})\right|^2\right], \label{NCEx}
\eea

where the expression for the Bogoliubov coefficients for the Rindler-Minkowski mode correspondence are given as

\bea 
\alpha_{\Omega \omega} =\frac{1}{2\pi a} \sqrt{\frac{\Omega}{\omega}}\exp{\left[\frac{\pi\Omega}{2a}\right]}\exp{\left[-\frac{i\Omega }{a} \log{\frac{\omega}{a}} \right]}\Gamma\left[\frac{i\Omega }{a}\right],\\
\beta_{\Omega \omega} =-\frac{1}{2\pi a} \sqrt{\frac{\Omega}{\omega}}\exp{\left[-\frac{\pi\Omega}{2a}\right]}\exp{\left[-\frac{i\Omega }{a} \log{\frac{\omega}{a}} \right]}\Gamma\left[\frac{i\Omega }{a}\right]. \label{BT}
\eea

Similar expression can be obtained for higher excited states. In \cite{AudretschMuller}, the authors evaluate the particle content for wave-packet basis in the inertial frame using states in Fock space, which will be a special construction of the general excited state (discussed over the appendices B, C, D) considered in this paper.

We can now evaluate each of the terms in the right hand side separately. Using the expressions in \ref{BT} we write

\bea
\int_0^{\infty} \frac{d \tilde{\omega}'}{\sqrt{4\pi\tilde{\omega}'}}\alpha^*_{\Omega \tilde{\omega}'}f(\tilde{\omega}')=\frac{1}{2\pi a} \exp{\left[\frac{\pi\Omega}{2a}\right]}\Gamma\left[-\frac{i\Omega }{a}\right]
\int_0^{\infty}\frac{d \tilde{\omega}'}{\sqrt{4\pi\tilde{\omega}'}}f(\tilde{\omega}')
\sqrt{\frac{\Omega}{\tilde{\omega}'}}\exp{\left[\frac{i\Omega }{a} \log{\frac{\tilde{\omega}'}{a}} \right]}. \label{FourierTransform1}
\eea

Now, making a transformation $\log{\frac{\tilde{\omega}'}{a}}=-t, $ we get

\bea
\int_0^{\infty}\frac{d \tilde{\omega}'}{\sqrt{\tilde{\omega}'}}\alpha^*_{\Omega \tilde{\omega}'}f(\tilde{\omega}')=\frac{\sqrt{\Omega}}{2\pi a} \exp{\left[\frac{\pi\Omega}{2a}\right]}\Gamma\left[-\frac{i\Omega }{a}\right]
\int_{-\infty}^{\infty} \frac{d t}{\sqrt{2\pi}} f(a e^{-t})\exp{\left[-\frac{i\Omega }{a}t \right]}. \label{FourierTransform2A}
\eea

Thus, we realize that the Fourier transform of the function $f(a e^{-t})/ \sqrt{2\pi}$ gives the dependence on $\Omega/a$ for the correction term of the number expectation value.

 We know that, when viewed from the Rindler frame, the frequency of the modes undergo a redshift \cite{PaddyBook} which involves the factor $e^{-t}$. Our result in \ref{FourierTransform2A} shows that the relevant expression is essentially a Fourier transform of a probability amplitude which is ``red-shifted''.  It is interesting to see that this feature is carried into a

superposed state as well.

Further, the distribution $f(\omega)$ gives the analytical properties of $N_{\Omega}$. We can verify, using the asymptotic form of the Gamma function, that prefactor in the above expression has the behavior:

\bea
\frac{\sqrt{\Omega}}{2\pi a} \exp{\left[\frac{\pi\Omega}{2a}\right]}\Gamma\left[-\frac{i\Omega }{a}\right] \rightarrow \text{constant}
\eea

for large $\Omega/a$. Therefore, if we wish the highest frequency modes to be unperturbed by the 

selection of the inertial state  we will require the Fourier transform to be decaying function in $\Omega/a$. (That is, if we want to choose  $f(\omega)$  such that the lower frequency modes are populated more effectively, which in turn will ensure that the excitation occurs beyond the length scale $1/a$ of the Rindler observer). We are interested in determining the condition for this behavior.

Let us call  the function $f(a e^{-t})/\sqrt{2\pi}$ as a new function $g(t)$. Then the normalizability of the state gives

\bea
\int_{-\infty}^{\infty}|g(t)|^2dt  < \infty,
\eea

showing that the function $g(t)$ is square integrable over $\mathbb{R}$. 

Using the 

properties of the Fourier transform of a square integrable function, we see that its Fourier transform must also be decaying\footnote{Fourier transform is a map from $\mathbb{L}^2$ to $\mathbb{L}^2$, which decay for large argument values.} at large 

$\Omega/a$ faster than at least $(\Omega/a)^{-1}$. In addition, if the function $g(t)$ is also absolutely integrable with its at least first $n$ derivatives also being so, then 

\bea
\left(-i\frac{\Omega}{a}\right)^n {\cal F}\{g\}=\int_{-\infty}^{\infty} d t g^{(n)}e^{\left[-\frac{i\Omega }{a}t \right]}, \label{Decay}
\eea

where ${\cal F}\{g\}$ is the Fourier transform of $g$. By virtue of $ g^{(n)}$ being an $\mathbb{L}^1$ function, the right hand side of \ref{Decay} is bounded. Thus

$|{\cal F}\{g\} | $ decays at least as fast as $(\Omega/a)^{-n} $.

For the other term 

\bea
\int_0^{\infty} \frac{d \tilde{\omega}}{\sqrt{4\pi\tilde{\omega}}}\beta_{\Omega \tilde{\omega}}f(\tilde{\omega}) 
\eea

in \ref{NCEx}, we realize the above mentioned criterion for $g$ and hence $f$ is suffice since the prefactor of the resultant Fourier transform

\bea
\frac{\sqrt{\Omega}}{2\pi a} \exp{\left[-\frac{\pi\Omega}{2a}\right]}\Gamma\left[\frac{i\Omega }{a}\right] \rightarrow 0
\eea

is anyway decaying for large $\Omega/a$ and will require only $g$ being absolutely integrable or that the Fourier transform of $g$ exists, which is true for square integrable functions. Thus, square integrable or physically meaningful states have decaying correction term to the number expectation value. We discuss some specific examples below.

\subsection{Example  I. $f(\omega)=C \sqrt{4 \pi \omega}\delta(\omega-\omega_0)$}

This case corresponds to the scenario, where the state is a 1-particle state of definite frequency. The constant $C$ has the mass dimension $1/2$ so as to make $f(\omega)$ dimensionless as required for $(1+1)$ dimensional space-time.

 Such a state will be useful in mapping the response of a definite momentum state as well. A definite momentum state is identified as a particle in the

quantum field theory and has been traditionally studied in details for high energy collision experiments. Thus we obtain the Rindler response to a single particle of the scalar field $\phi$. 

In this case

\bea
g(t)=C (a e^{t})^{1/2}\delta(a e^t-a e^{t(\omega_0)}).
\eea

Since the Dirac delta function is nether square integrable nor absolutely integrable in $(-\infty, \infty)$ we expect the correction term \textit{not} to be decaying asymptotically. One can work out 

the expressions in \ref{NCEx} to verify that

\bea 
N_{\Omega}=\frac{|C|^2}{2\pi a \omega_0} \coth{\left(\frac{\pi\Omega}{a}\right)}, \label{1ptNCorr}
\eea

which saturates asymptotically. The effect of non-vacuum state can be felt even for the largest modes of the Rindler observer in this case. This can be checked even for the $n-$particle definite momentum sates or their superpositions.

This is an important and curious result. It tells us that an one particle state in Minkowski frame will lead to the correction term (over and above the thermal contribution) to the expectation value of the number operator. In particular, this correction persists at all frequencies and will dominate over the thermal component at high frequencies.  The correction term in this case keeps the memory of the initial excitation in the inertial frame through $\omega_0$. The exact profile of the correction depends on both the initial excitation mode $\omega_0$ and the mode $\Omega$ at which the observation is made. (This result has significant implication for the corresponding black hole scenario.  We hope to study the implications of this result for the black hole in a future publication.) 

However, we also note that the state considered here is not a physical one in the sense that we can not prepare the state with arbitrary precision about the momentum. In other words, such a state is not normalizable. For, physically realistic cases we need to consider only those states which are square integrable. We discuss two such examples below.

\subsection{Example II. $f(\omega)=C\omega\exp{[-\alpha \omega ]}$}

This is a case where we decide to excite the  modes and the width and mean value of the excitation controlled by a parameter $\alpha$. In this case, the constant $C$ has the mass dimension $-1$ and we have :

\bea
g(t)=Cae^t \exp{[-\alpha a e^t]}.
\eea

We see that this function is square integrable and all its derivatives are absolutely integrable too, thus the corrected number expectation value should decay very fast asymptotically. Indeed, the evaluation for \ref{NCEx} gives

\bea 
N_{\Omega}=\frac{|C|^2}{2 a \alpha^2}\frac{\coth{\left(\frac{\pi\Omega}{a}\right)}}{\sinh{\left(\frac{\pi\Omega}{a}\right)}}
\eea

which clearly decays down for large $\Omega/a$. We realize that the correction term is a monotonically decreasing function in $\Omega$ with a pole at $\Omega=0$. The pole at the zero energy mode is second order in this case and supersedes the usual pole in the thermal contribution. Therefore, the dominant departures from thermality will be visible towards the zero energy modes, while the high energy probes in Rindler frame will receive progressively diminishing corrections.  The information about the excitation in the inertial modes is captured in the parameter $\alpha$ sitting in the denominator as before.

\subsection{Example III. $f(\omega)=C\sqrt{\omega}\frac{1}{\sqrt{2\pi \sigma^2}}\exp{[-\frac{1}{2\sigma^2}( \omega-\omega_0)^2]}$}

This is a Gaussian distribution peaked about a specified frequency. In this case, the constant $C$, again has the mass dimension $1/2$.

This distribution also dies down for large $\omega$ and hence large $k$-modes. The function $g(t)$ and all its derivatives are absolutely integrable in this case as well. Thus the correction to the number expectation value becomes

\bea 
N_{\Omega}=\frac{|C|^2}{8\sqrt{2}\pi^2a \sigma }\coth{\left(\frac{\pi\Omega}{a}\right)}\times|{\cal Z}(\Omega,\omega_0,a,\sigma)|^2, \label{GaussianCorr}
\eea

where

\bea
{\cal Z}(\Omega,\omega_0,a,\sigma)=\Gamma\left[\frac{a+2i \Omega}{4 a} \right]\tilde{F}_1^1\left(\frac{a-2i \Omega}{4 a},\frac{1}{2},-\frac{\omega_0^2}{2\sigma^2} \right)+\frac{\sqrt{2}\omega_0}{\sigma}\Gamma\left[\frac{3a+2i \Omega}{4 a} \right]\tilde{F}_1^1\left(\frac{3a-2i \Omega}{4 a},\frac{1}{2},-\frac{\omega_0^2}{2\sigma^2} \right). 
\eea

By appealing to the asymptotic properties of the Gamma function and the confluent Kummer Hypergeometric function $\tilde{F}_1^1\left(z,n,b\right) $ we can verify that the correction term indeed vanishes at large $\Omega/a$.  Further, exploiting the properties of the Kummer Hypergeometric functions for low $\sigma$ values, we can verify that the decaying part becomes constant in the $\sigma \rightarrow 0$ limit, leaving a $\coth{\left(\frac{\pi\Omega}{a}\right)} $ dependence in \ref{GaussianCorr}, as one would have expected (If negative frequency modes are also taken, the limit $\sigma \rightarrow 0$ sends the Gaussian distribution to a delta function). We can also verify that smaller and smaller values of $\sigma$ will allow the correction to be perceived at larger and larger $\Omega$ modes. The extreme case is that of the delta function, when correction saturates to a constant value for large $\Omega$. In other words, the result obtained in Example I above for a $1-$particle state in the Fock basis emerges as a sensible limiting case of a physically realizable state when we make the state more and more sharply defined in the momentum state.

Therefore, we see that for single excitation state, the correction term has decaying profile in $\Omega/a$ by virtue of  $g$ being an $\mathbb{L}^2$ function. Additionally, if the function $g$ happens to be an $\mathbb{L}^1$ function with at least some of its derivative also behaving the same way, the correction term will fall correspondingly faster. We have seen some distributions which defy that and have non vanishing correction at asymptotia. However those distributions are also not $\mathbb{L}^2$. The correction term is sum of two Fourier transforms of the function $g$, we realize now that it is asymptotically dying profile, since it is sum of magnitudes of two square integrable functions. A very similar exercise will establish corresponding results for an $n$-particle as well as a general superposed state, which we present in Appendix B and Appendix C respectively.

We will now discuss the response of the Unruh-DeWitt detector in a similar setting, where the inertial state is arbitrary. In that case, the two-point function is not time translation invariant anymore, as would have happened if the observer were moving on a trajectory which is not the integral curve of a time-like Killing field (even for the vacuum state).

\section{ Detector response for the non-vacuum Minkowskian state}

We have a precise understanding of the response of a Rindler detector when the field happens to be in the vacuum state of the Minkowski modes. In this case the two point correlation function happens to be time translation invariant. So, the transition \textit{rate} can just be viewed as the time translation invariant integrand in the expression of transition probability \ref{ProbVac}. The vacuum part then contributes a steady rate of excitation involving a thermal distribution in the energy.

Our interest is in  the cases where the initial state does not happen to be the Minkowski vacuum for different trajectories of detector. In those cases we will require the two point correlation function $\langle \Psi |\phi[x(\tau')]\phi[x(\tau)]|\Psi\rangle,$ for the initial state $|\Psi\rangle$.

Let us begin with an one particle state of a definite momentum in Minkowski frame with

$f(k)=C \sqrt{4\pi\omega_k}\delta(k-k_0)$,

where $k_0>0$ viewed by  Rindler observers with  the trajectory:

\bea
x(\tau)= a^{-1} \cosh\left(a \tau\right) \nonumber\\
t(\tau)= a^{-1} \sinh\left(a \tau\right), \label{RindlerTraj}
\eea

Using \ref{eff4mink} and \ref{RindlerTraj}, we find that the effective field is:

\bea
\Phi_{\text{eff}}(x)=\frac{C}{\sqrt{2\omega_{k_0}}}e^{i k_0\cdot x} = \frac{C}{\sqrt{2\omega_{k_0}}}e^{i \frac{k_0}{a}e^{-a\tau}}.
\eea

Equations \ref{1stOrderTransition} and \ref{RindlerTraj} now give us

 the correction for the finite time transition probability:

\bea
\bar{P} =\frac{|c|^2|\langle E|\hat{m}(0)|E_0\rangle |^2}{4\pi\omega_{k_0}a}\left[\left|\zeta\left(\frac{\Delta E}{a},\frac{k_0}{a}\right)\right|^2 + \left|\zeta\left(\frac{\Delta E}{a},-\frac{k_0}{a}\right)\right|^2 \right], \label{TransitionProb4delta}
\eea

with

\bea
\zeta\left(\frac{\Delta E}{a},\frac{k_0}{a}\right)=|C|^2\left[e^{i \Delta E T} Ei\left[1+\frac{i \Delta E}{a}, i\frac{k_0}{a} e^{-T/a} \right]-e^{i \Delta E T_0} Ei\left[1+\frac{i \Delta E}{a}, i\frac{k_0}{a} e^{-T_0/a} \right]  \right],
\eea

where $Ei[n,x]$ is Exponential integral function of order $n$.

The corresponding transition rate is obtained by taking a derivative of \ref{TransitionProb4delta} with respect to $T$. 

An important point to note is that both the finite time probability of transition and the rate  have now become $T-$dependent. Further, the correction part for the rate can (in principle) turn negative for some particular choices of the parameters involved. If the correction term is large then at some scales the total transition rate (including the thermal part) can also turn negative. (A plot of the transition rate  confirms the oscillatory feature.)

In those cases in which  the correlation function depends only on $\tau-\tau'$, another formal definition of the rate can be given, which is often used in the literature. This is given, essentially by dropping one of the (infinite) time integrations:

\bea
\bar{R} =|c|^2\int_{-\infty}^{\infty}d(\tau-\tau') e^{i\Delta E(\tau-\tau')}|\langle E|\hat{m}(0)|E_0\rangle |^2 
\langle \Psi |\phi[x(\tau')]\phi[x(\tau)]|\Psi \rangle. \label{Rate1}
\eea

In this case the expressions for the rate in \ref{Rate1} coincides with the derivative of \ref{TransitionProb4delta}. However, in generic cases where the two-point function will not have the time translation invariance, transition rates obtained from these two different routes will be different. We will discuss this aspect in more detail in the next section.

To obtain the expression for the rate in this manner, for the definite momentum state, we evaluate the probability once again, in terms of the variables

$T=(\tau + \tau')/2$ and
$s= \tau -\tau'$.

The two-point function is 

\bea
\langle k_0 | \phi(x)\phi(y) | k_0 \rangle ={}_M\langle 0| \phi(x)\phi(y) | 0 \rangle_M +\frac{|C|^2}{4\pi\omega_{k_0}}e^{-i k_0\cdot (x-y)}+\frac{|C|^2}{4\pi\omega_{k_0}}e^{i k_0\cdot (x-y)}. \label{2pt41pt}
\eea

In terms of the new variables, we have the expression:

\bea
e^{-i k_0\cdot (x-y)}=e^{+i \frac{k_0}{a} e^{-a T}(e^{-a s/2}-e^{a s/2})}. \label{expressionRindler}
\eea

Using \ref{2pt41pt} and \ref{expressionRindler}, we can write the transition probability as

\bea
\bar{P} =|c|^2|\langle E|\hat{m}(0)|E_0\rangle |^2 \int_{-\infty}^{\infty} dT \int_{-\infty}^{\infty} ds e^{-i\Delta E s} \times \nonumber\\
\left[{}_M\langle 0| \phi(x)\phi(y) | 0 \rangle_M + \frac{|C|^2}{4\pi\omega_{k_0}} e^{+i \frac{k_0}{a} e^{-a T}(e^{-a s/2}-e^{as/2})} + 
\frac{|C|^2}{4\pi\omega_{k_0}} e^{-i \frac{k_0}{a} e^{-a T}(e^{-a s/2}-e^{as/2})} \right].
\eea

Henceforth, we will drop the pre-factors $|c|^2|\langle E|\hat{m}(0)|E_0\rangle |^2 $, in a sense that probability will be considered in its units.

Doing a variable transformation 

$e^{-a s/2} = z$, we obtain the correction over the usual vacuum thermal part to be:

\bea
P =\int_{-\infty}^{\infty} dT \frac{|C|^2}{4\pi\omega_{k_0}}\left[\frac{2}{a}\int_0^{\infty} dz z^{2i\frac{\Delta E}{a}-1}
\left(e^{+i \frac{k_0}{a} e^{-a T}\left(z-\frac{1}{z}\right)}+e^{-i \frac{k_0}{a} e^{-a T}\left(z-\frac{1}{z}\right)} \right) \right],\\
=\frac{|C|^2}{a \pi\omega_{k_0}} \int_{-\infty}^{\infty} dT  \left[K_{2i\frac{\Delta E}{a}}\left(2\frac{k_0}{a} e^{-a T}\right)e^{-\frac{\pi}{a}\Delta E}+ K_{-2i\frac{\Delta E}{a}}\left(2\frac{k_0}{a} e^{-a T}\right)e^{\frac{\pi}{a}\Delta E}\right],\label{prob41ptA}\\
=\frac{|C|^2}{4 a^2 \pi\omega_{k_0}}\left[(e^{-\frac{\pi}{a}\Delta E}+ e^{\frac{\pi}{a}\Delta E}) \Gamma\left[-\frac{i \Delta E}{a}\right] \Gamma\left[\frac{i \Delta E}{a} \right]\right]
= \frac{|C|^2}{2 a \Delta E \omega_{k_0}}\coth{\left(\frac{\pi \Delta E}{a}\right)}.
\label{Prob41pt}
\eea

This correction term has a simple interpretation and corresponds to the number expectation value in this state obtained earlier in \ref{1ptNCorr} (There is a factor $\pi$ discrepancy though which is discussed in the Appendix D).

In fact, this correspondence holds even for a general single excitation state (and a most general state) as we shall see later. 

It is also interesting to note that at energies higher than the Rindler scale the correction saturates to $|C|^2/2 a \Delta E \omega_{k_0}$.

From the expression for the total probability for transition  in \ref{prob41ptA}, we can identify a transition \textit{rate} as being given by the integrand of the integral occurring in \ref{prob41ptA}. Adding the rate contributed by the vacuum state (which is not included in \ref{prob41ptA}, which computes the \textit{corrections} to the vacuum result), we get: 

\bea 
\bar{R} =\frac{1}{\Delta E}\frac{1}{e^{\frac{2\pi}{a} \Delta E }-1}+\frac{|C|^2}{a \pi \omega_{k_0}} \left[K_{2i\frac{\Delta E}{a}}\left(2\frac{k_0}{a} e^{-a T}\right)e^{-\frac{\pi}{a}\Delta E}+ K_{-2i\frac{\Delta E}{a}}\left(2\frac{k_0}{a} e^{-a T}\right)e^{\frac{\pi}{a}\Delta E}\right].
\eea

Clearly this transition rate  is time dependent (i.e depends on $T$) and  does not necessarily always remain positive definite. (A plot of the rate illustrates the oscillatory nature.) It will also become negative at various values of $\Delta E$ and $T$.

We will address the properties of the transition rate in more detail in the next section.

Since the quantum state is not a vacuum state even an inertial detector will also detect particles in this state. At any given propertime  $\tau$, one can introduce an inertial detector which is co-moving with our accelerated detector, in the sense that it will have the same velocity as the non-inertial detector at $\tau$. Such a  co-moving inertial detector  at  time  $\tau$ will observe a transition rate: 

\bea
R_{\text{inertial}} = \frac{|C|^2}{2 \omega_{k_0}} \delta(\Delta E - k_0 \gamma(1-v)),
\eea

where $\gamma$ is the Lorentz factor for the observer moving at the velocity $v = \tanh{(a \tau)}$ and $k_0 >0$. (This expression can be easily obtained from the detailed discussion on transition rate presented in the next section.) The total transition registered by a set of co-moving inertial observers distributed along the trajectory of the Rindler observer, is therefore given by:

\bea
P_{\text{inertial}}=\int_{-\infty}^{\infty} d \tau R_{\text{inertial}}[v(\tau)] = \frac{|C|^2}{2 a \Delta E \omega_{k_0}}.
\eea 

This is similar to the asymptotic limit of the expression for $N_\Omega$ in ......\marginpar{fill Eq No} obtained earlier.  This is also the $a\rightarrow 0$ limit of \ref{prob41ptA} as also argued in \cite{PaddyTP} for inertial observers' computation.

\subsubsection{Normalizable states}

We will next consider the detector response for  general singly excited state which is normalizable (unlike the delta function state considered earlier). The state \ref{1PS}, in $1+1$ dimension becomes

\bea
|1\rangle_M = \int_{-\infty}^{\infty} \frac{dk}{\sqrt{2\pi}}\frac{f(k)}{\sqrt{2\omega_k}}a^{\dagger}(k)|0 \rangle_M.
\eea

As discussed previously, the effective field for this state is given as

\bea
\Phi_{\text{eff}}=\int_{-\infty}^{\infty} \frac{dk}{2\pi}\frac{f(k)}{2\omega_k} e^{i k \cdot x}.
\eea

Now, if we use the Rindler co-ordinates \ref{RindlerTraj} and identify 

\bea
\left(\frac{k}{a}\right)^{-i\frac{\Delta E}{a}}= e^{-i\frac{\Delta E}{a}\log{\left(\frac{k}{a}\right)}},
\eea

and make the transformation to a new variable  $\log{\left(\frac{k}{a}\right)}=t$, all the integrals involved in the computation get converted into the Fourier transforms of square integrable functions in $t$, with respect to $\Delta E/a$ (see Appendix E). Therefore, the correction, as observed by the Rindler detector decays  for large 
$\Delta E/a$.
 This behavior was earlier observed in the number expectation value for Rindler observer as well. We note that, had the two point correlation function been time translation invariant in detector's co-ordinates, we would have been able to define easily, the probability of transition as a function of number (density) expectation value.

For computing the correction in the number expectation value for the Rindler observer, we confined ourselves to one of the Rindler wedges and used only the left-moving modes. However, in general, there will be contributions from both the left and right wedge  as well as  from the left-moving and the right-moving modes. For the sake of completeness, we show in Appendix E that, 
taking into account of all these terms does not change the result.

A very similar calculation will yield the same result for arbitrary superposed states. Since the $x$- dependence in the transition amplitude is brought in only through the field operators (in the correlation function) and is independent of the state, the $s$- and $T$- integrals that are to be performed remain the same. The only difference will be the introduction of many $\hat{a}_k$s and $\hat{a}_{k'}^{\dagger}$s squeezed between the Minkowskian vacuum state, in addition to the introduction of many 
distribution functions $f(k)$s as well. In a fashion exactly as before (when we  evaluated the number expectation value), such terms will give a contribution like,
\bea
C^{ij}f_i(k)f_j(k'),
\eea
(see Appendix B),
and using the $s$- and $T$- integrals we can again show that the correction term is a superposition of products of the Fourier transform of various
square integrable functions with respect to $\Delta E/a$. Hence, exactly as before, the detector sees a correction which  decays for large
$\Delta E/a$ for any arbitrary but physical (normalizable) state. These results have significant implications in their own right. We see that sufficiently high energy probes progressively forget about the distribution function in inertial modes (or about the inertial state itself). These observers find themselves in a thermal bath which does not depend in any fashion on the details of the initial inertial state, but one decided by the vacuum structure of the theory. This evidently will have significant implications for black hole thermodynamics. For example, if a black hole made up of by collapse of a field in a non-vacuum state, any UV-probe of the black hole will only reveal the thermal character at the Hawking temperature.

We will next consider the rate of transitions of the detectors in detail in the next section. In this case we see that the rate is time dependent and different definitions will give rise to different expressions for the time dependent transition rate. We will also note some important properties of the transition rates.

\section{Rate of transition}

As we discussed in the previous section, one way of defining the transition rate from the probability of transition is by using \ref{Rate1}. If the two-point function of the field is time translation invariant (which is the case for the vacuum state), then rate \ref{Rate1} is constant and is independent of  $T$. 
However, in general time translation invariance will not hold for arbitrary superposed states. Therefore  we see from \ref{prob41ptA} (and \ref{integration2}-\ref{integration5A} in Appendix E), that the transition rate defined by this procedure will become $T$-dependent.

There is an alternative way of defining the detector rate which is frequently used in the literature. This is done by keeping the detector ``on'' for a finite duration, say $(0,T)$ and define the transition rate $R$ as $R=dP/dT$ where $P(T)$ is the probability for transition. Such a definition, used extensively in literature \cite{Rate}, shows some peculiar features for different states \cite{Luoko2} possibly due to effect of switching off at finite time sharply.

We will now analyze some  properties of the rates defined by these two procedures. We first discuss the ``usual'' method of defining the rate as the integrand of the $T-$ integral and then discuss the second case.

\subsection{Transition Rate I}

\subsubsection{Transition rate as the Wigner function}

From \ref{TwoPointFunction} we see that the additional contribution (apart form the standard vacuum term)  to the correlation function is given by:
\bea
{\cal C}(x,y)= \int_{-\infty}^{\infty} \frac{dk}{2\pi}\int_{-\infty}^{\infty} \frac{dk'}{2\pi}\frac{f(k)}{2\omega_k}\frac{f^*(k')}{2\omega_k'}e^{-ik\cdot x+ik'\cdot y} +\text{c.c.} \label{2ptFcorr}
\eea

As discussed previously, can rewrite the above terms using $x\equiv x[\tau]$ and $y\equiv y[\tau]$  as 
\bea
{\cal C}(\tau,\tau')=2\text{Re}[\Phi_{\text{eff}}(\tau) \Phi_{\text{eff}}^*(\tau')],\label{2ptFcorr}
\eea

with
\bea 
\Phi_{\text{eff}}(\tau)=\int_{-\infty}^{\infty} \frac{dk}{2\pi}\frac{f(k)}{2\omega_k}e^{ik\cdot x[\tau]}=\int_{-\infty}^{\infty} \frac{dk}{2\pi}\frac{f(k)}{|k|}e^{-i|k|t(\tau) +ik x(\tau)}.
\eea

We can also rewrite $\Phi_{\text{eff}}(\tau)$ in frequency modes  with the help of ${\cal F}_{\omega}$ as
\bea
\Phi_{\text{eff}}(\tau)=\int_{-\infty}^{\infty} \frac{d\omega}{2\pi} {\cal F}_{\omega} e^{-i\omega \tau}.
\eea
Therefore, the correction term \ref{2ptFcorr} becomes
\bea  
\int_{-\infty}^{\infty} \int_{-\infty}^{\infty} \frac{d\omega}{2\pi}\frac{d\omega'}{2\pi}{\cal F}_{\omega}{\cal F}_{\omega'}^* e^{-i\omega \tau +i\omega' \tau'} +\text{c.c.}
\eea
From this correction term we can calculate the correction to the transition rate over the standard thermal contribution. The correction due to the first term is: 
\bea
\int_{-\infty}^{\infty} ds e^{-i\Delta E s}\int_{-\infty}^{\infty} \int_{-\infty}^{\infty} \frac{d\omega}{2\pi}\frac{d\omega'}{2\pi}{\cal F}_{\omega}{\cal F}_{\omega'}^* e^{-i\omega\left(T+\frac{s}{2}\right)+ i\omega'\left(T-\frac{s}{2}\right)}&=&\nonumber\\
\int_{-\infty}^{\infty} \int_{-\infty}^{\infty} \frac{d\omega}{2\pi}\frac{d\omega'}{2\pi}{\cal F}_{\omega}{\cal F}_{\omega'}^*2\pi\delta\left(\frac{\omega+\omega'}{2}-\Delta E \right)e^{-i(\omega-\omega')T} 
&=& 2\int_{-\infty}^{\infty} \frac{d\omega}{2\pi}{\cal F}_{\omega}{\cal F}_{2\Delta E-\omega}^*e^{-2i(\omega-\Delta E)T}.
\eea
Making another transformation
$\tilde{\omega}=\omega-\Delta E$, the above term becomes
\bea
2\int_{-\infty}^{\infty} \frac{d\tilde{\omega}}{2\pi}{\cal F}_{\Delta E-\tilde{\omega}}^*e^{-2i\tilde{\omega}T}{\cal F}_{\Delta E+\tilde{\omega}} ={\cal W}_{\cal F}(\Delta E, T),
\eea
where ${\cal W}_{\cal F}$ is the standard Wigner function (defined in the time-frequency domain) corresponding to the function ${\cal F}$. It is easy to see that the second term will contribute the same term as the correction. Therefore the net correction term is twice the Wigner function of ${\cal F}$. 

The properties of the function ${\cal F},$ which is the Fourier counterpart of the effective field $\Phi$, will evidently decide the properties of the correction to the rate of the transition.  We know that the Wigner function is well defined and bounded for square integrable functions. Thus, the behavior of the correction will depend upon whether ${\cal F}$ is square integrable or not. Further, the Wigner function is positive definite only for Gaussian functions and in general  it can take negative values. However, for the total transition rate to turn negative we need the negative value of the Wigner function to be larger in magnitude than the vacuum transition rate part. 

Therefore, unlike the vacuum state the transition rate for a non-vacuum state remains non-constant and oscillatory in nature. Thus, the cases where the total transition rate becomes negative, there will be transitions occurring from the excited state of the detector to the ground state. The total number of particles in the excited state of the detector can in principle go down in time too. Therefore, in principle such states can give rise to de-excitations in the detector as well. Such a scenario is also discussed in \cite{AudretschMuller} for the wavepacket basis description of the field.

\subsubsection{Correction term for Minkowski detector}

A detector at rest (or moving with uniform velocity) will not click in the inertial vacuum sate. But if the field is in an excited state (like the one particle state we are studying), a detector at rest can get excited by absorbing the quanta of the field. it is useful to start our discussion by studying the case of a uniform velocity detector, in order to delineate these effects.
For such a detector

\bea
t(\tau)=vx_0 + \gamma \tau \nonumber\\
{\bf x}(\tau)=x_0+v\gamma \tau  \label{MinkowskiTraj}
\eea

Therefore, 

\bea
\Phi_{\text{eff}}(\tau) = \int_{-\infty}^{\infty} \frac{dk}{2\pi}\frac{f(k)}{2\omega_k}e^{-ik\cdot x(\tau)}=\int_{-\infty}^{\infty} \frac{dk}{2\pi}\frac{f(k)}{2|k|}e^{-i|k|t(\tau)+ik x(\tau)}, \nonumber\\
          = \int_{-\infty}^{\infty} \frac{dk}{2\pi}\frac{f(k)}{2|k|}\left(e^{-i|k|vx_0 +i k x_0}\right)e^{-i(|k|-kv)\gamma t(\tau)}.
\eea
Again, we can obtain the expression for ${\cal F}$ ,
\bea
{\cal F}(\omega) = \int_{-\infty}^{\infty} \Phi_{\text{eff}}(\tau) e^{i \omega \tau} d\tau 
= \int_{-\infty}^{\infty} \frac{dk}{2\pi}\frac{f(k)}{2|k|}\left(e^{-i|k|vx_0 +i k x_0}\right)\int e^{-i(|k|-kv)\gamma t(\tau)+i \omega \tau}d\tau \nonumber\\
=\int_{-\infty}^{\infty} dk \frac{f(k)}{2|k|}\left(e^{-i|k|vx_0 +i k x_0}\right)\delta(\omega-(|k|-kv)\gamma).
\eea
Writing the delta function in terms of its roots as
\bea
\delta\left(\omega-(|k|-kv)\gamma\right)=\frac{\delta\left(\frac{\omega}{(1-v)\gamma}-k\right)}{(1-v)\gamma}+\frac{\delta\left(\frac{\omega}{(1+v)\gamma} + k\right)}{(1+v)\gamma},
\eea
we get the result:
\bea
{\cal F}(\omega)=\left[\frac{f\left({\cal D}\omega\right)}{2\pi|\omega|} e^{-i\omega x_0/\gamma}+\frac{f\left(\frac{-\omega}{{\cal D}}\right)}{2\pi|\omega|} e^{i\omega x_0/\gamma}  \right], \label{1ptDoppler}
\eea
with 
\bea
{\cal D} = \sqrt{\frac{1+v}{1-v}},
\eea
being the Doppler shift factor. So, this expression has the simple interpretation in terms of the Doppler shifted frequency in the moving frame. For a trivial example we consider the case for the rest frame, i.e. $v=0$, $\gamma=1$. In this case, the expression \ref{1ptDoppler} becomes
\bea
{\cal F}(\omega)=\left[\frac{f\left(\omega\right)}{2\pi|\omega|} e^{-i\omega x_0}+\frac{f\left(-\omega \right)}{2\pi|\omega|} e^{i\omega x_0}  \right],
\eea
which just picks the contributions from both positive as well as the  negative frequency modes.

We note that for a physically realistic state $f(k)/{\sqrt{2|k|}}$ is square integrable, but that does not guarantee the same for ${\cal F}(\omega)$,  which includes an additional $\sqrt{\omega}$ contribution in the denominator making it divergent at small $\omega$ values. 
Therefore, the Wigner function corresponding to ${\cal F}$ might lack the nice feature of boundedness. Therefore, a very large correction has the potential to supersede the thermal part in the full transition rate. Hence, the transition rate defined in the fashion discussed above has the potential to turn negative even in the inertial frame. We have already seen such a case for the Rindler observers too. 

\subsubsection{Correction term for the Rindler observer}

For Rindler observer  the effective field is given by
\bea
\Phi_{\text{eff}}(\tau)=\int_0^{\infty} \frac{dk}{2\pi}\frac{f(k)}{2k}e^{i\frac{k}{a}e^{-a\tau}}+\int_0^{\infty} \frac{dk}{2\pi}\frac{f(-k)}{2k}e^{-i\frac{k}{a}e^{a\tau}}.
\eea

Therefore, 

\bea
{\cal F}(\omega)=\int_0^{\infty} \frac{dk}{2\pi}\frac{f(k)}{2k}\int_{-\infty}^{\infty} d\tau e^{i\frac{k}{a}e^{-a\tau}+i\omega\tau}+\int_0^{\infty} \frac{dk}{2\pi}\frac{f(-k)}{2k}\int_{-\infty}^{\infty} d\tau e^{i\frac{k}{a}e^{-a\tau}+i\omega\tau}.
\eea
Using
\bea
\int_{-\infty}^{\infty} d\tau e^{i\frac{k}{a}e^{-a\tau}+i\omega\tau}=\frac{1}{a}\left(-i\frac{k}{a}\right)^{i\omega/a}\Gamma\left[-i\frac{\omega}{a}\right], \label{tauintegral1}
\eea
we obtain,
\bea
{\cal F}(\omega)=\frac{e^{\frac{\pi \omega}{2 a}}}{2\pi a}\left( \Gamma\left[-i\frac{\omega}{a}\right]+  \Gamma\left[i\frac{\omega}{a}\right]\right)\int_0^{\infty}\frac{dk}{2\pi}\frac{f(k)}{2k}\left(\frac{k}{a}\right)^{i\omega/a}.
\eea

As before the integral is a Fourier transform of an $\mathbb{L}^2$ function. It is  multiplied with a regular function  which decays  for large values of $\omega$. Thus the function ${\cal F}(\omega)$ itself is square integrable and the correction term, as seen by the Rindler detector, remains finite. The total transition \textit{rate} in this case remains positive only as long as the magnitude of the correction part is smaller than the standard vacuum transition rate.

\subsection{Transition Rate II}

As discussed earlier, there exists another method to define the transition rate by obtaining the transition \textit{probability} up to some finite time and then take the time derivative. We will first  obtain the relation between these two approaches. For a simple demonstration, we will consider quantum states in which only  positive $k$ modes are excited (only left moving ones, say). (In that case only the contribution from the first two terms in \ref{CorwithDot} in Appendix E, will survive. In principle, all other terms can also be considered within similar set up). In order to compare the  rates defined by the two procedures, we need to obtain the expression for the rate in \ref{RateFormal} in late time limit. This is because because the second procedure  to define the rate --- using the integrand of $T$ integration --- requires the limits of $\tau$ and $\tau'$ to run from $(-\infty,\infty)$.
The time-dependent rate for such a configuration, defined as the integrand of $T$, is given by:
\bea
R _1(\Delta E)|_{T\rightarrow \infty}=\frac{ 4 e^{-\pi \Delta E/a}}{ a} \int \frac{dk}{2\pi}\int \frac{dk'}{2\pi}\frac{f(k)}{2\omega_k}\frac{f^*(k')}{2\omega_k'} \left(\frac{k}{a}\right)^{-i\Delta E/a} \left(\frac{k'}{a}\right)^{+i\Delta E/a} K_{-2i\Delta E/a}\left[ 2\sqrt{\frac{k}{a} \frac{k'}{a}} e^{-aT}\right]_{T\rightarrow \infty}, \label{TransitionRate1}
\eea
while the expression for the rate defined formally in \ref{RateFormal} is:
\bea
R _2(\Delta E)|_{T\rightarrow \infty}=\frac{ e^{-\pi \Delta E/2a}}{a}\int \frac{dk}{2\pi}\int \frac{dk'}{2\pi}\frac{f(k)}{2\omega_k}\frac{f^*(k')}{2\omega_k'} \left[ \left(\frac{k e^{-aT}}{a}\right)^{-i\Delta E/a} \Gamma\left[i\frac{\Delta E}{a}\right] +
\left(\frac{k' e^{-aT}}{a}\right)^{i\Delta E/a}\Gamma\left[-i\frac{\Delta E}{a}\right]\right]_{T\rightarrow \infty} \label{TransitionRate2}
\eea
clearly showing that the two approaches for defining the transition rates \textit{do not agree} when the correlation function lacks the property of time translation invariance in $\tau$. In fact this is a \textit{purely algebraic result} and it can be demonstrated that, for an arbitrary function of two variables,  the rates defined by the two different procedures will result in two different expressions.
To see this consider:
\bea
I_1 = \int \int e ^{-i \Delta E (t_1-t_2)} f(t_1,t_2) dt_1 dt_2. \label{I_1}
\eea
and write the function $ f(t_1,t_2) $ in terms of the Fourier modes
\bea
f(t_1,t_2)=\frac{1}{4 \pi^2} \int _{-\infty}^{\infty}\int_{-\infty}^{\infty} d \kappa d \xi \hat{f}(\kappa, \xi) e^{i \kappa t_1+ i \xi  t_2}, 
\eea
leading to:
\bea
I_1 = \frac{1}{4 \pi^2} \int_{-\infty}^{\infty} \int_{-\infty}^{\infty} d \kappa d \xi \left[\int dt_1 e^{i (\kappa-\Delta E) t_1} \int dt_2 e^{i (\xi +\Delta E) t_2} \right] \hat{f}(\kappa, \xi)
\eea
Now if the $t_1, t_2$ integrals run from $(-\infty,\infty)$, we simply get 
\bea
I_1= \hat{f} (\Delta E,-\Delta E).
\eea
If the integrals run form $(-T, T)$ we have
\bea
I_1 = \frac{1}{4 \pi^2} \int_{-\infty}^{\infty} \int_{-\infty}^{\infty} d \kappa d \xi \left[\int_{-T}^{T} dt_1 e^{i (\kappa-\Delta E)} t_1 \int_{-T}^{T} dt_2 e^{i (\xi +\Delta E)} t_2 \right] \hat{f}(\kappa, \xi).
\eea
In this case the rate, calculated with respect to $T$ is:
\bea
\frac{dI_1}{dT} = \frac{1}{4 \pi^2}\left[\int_{-\infty}^{\infty} \int_{-\infty}^{\infty} d \kappa d \xi \hat{f}(\kappa, \xi) e^{i (\kappa-\Delta E)T} \int_{-T}^{T} dt_2 e^{i (\xi +\Delta E)} t_2+ e^{i (\xi +\Delta E)T} \int_{-T}^{T} dt_1 e^{i (\kappa-\Delta E)} t_1 \right].
\eea
In the late time limit $T\rightarrow \infty$, the  expression for the rate becomes
\bea
\lim_{T\rightarrow\infty}\frac{dI_1}{dT}  = \lim_{T\rightarrow\infty}\left[\frac{e^{-i \Delta E T}}{2 \pi}  \int_{-\infty}^{\infty} d \kappa\hat{f}(\kappa, -\Delta E) e^{i \kappa T} +
\frac{e^{i \Delta E T}}{2 \pi}  \int_{-\infty}^{\infty} d \xi \hat{f}(\Delta E, \xi) e^{i \xi T}\right]. \label{Rate1Expression}
\eea
We now change variables to $ {\cal T}=(t_1+t_2)/2$ and
$s= t_1-t_2$ which run from ${\cal T}\in(-T,T)$ and $s\in(-2T,2T)$; then the rate, defined as $dI_1/d{\cal T} $, is given by:
\bea
\frac{dI_1}{d{\cal T}} = \frac{1}{4 \pi^2}  \int_{-\infty}^{\infty} \int_{-\infty}^{\infty} d \kappa d \xi \hat{f}(\kappa, \xi)e^{i (\kappa +\xi ){\cal T}}\int_{-2T}^{2T} d s e^{i\left(\frac{\kappa-\xi}{2}-\Delta E  \right) s},
\eea
which for the late time limit becomes
\bea
\lim_{{\cal T}\rightarrow\infty}\frac{dI_1}{d{\cal T}} =\lim_{{\cal T}\rightarrow\infty}\left[\frac{e^{-2 i\Delta E {\cal T}}}{\pi}\int_{-\infty}^{\infty} d \kappa\hat{f}(\kappa, \kappa -2 \Delta E )e^{-2 i \kappa {\cal T}}\right]. \label{Rate2Expression}
\eea
It is now clear, on comparing \ref{Rate1Expression} and \ref{Rate2Expression} that the two approaches of defining the transition rate are different in general. At late time limit the two expressions can be written in terms of Fourier transform of a bivariate function. In this light, we can consider the rates for the Rindler observer. The Rindler trajectory discussed earlier corresponds to 
\bea
f(t_1,t_2)=e^{i(\frac{k}{a}e^{-at_1}-\frac{k'}{a}e^{-at_2})},
\eea
for which we obtain the expressions for the rates at late times as shown in \ref{TransitionRate1} and \ref{TransitionRate2}, respectively. The transition rate defined by either procedure remains real though not necessarily positive definite. (The reality condition is also evident from \ref{RateFormal}). We see from \ref{TransitionRate2} that even for finite time operation of the detector (when it loses the nice correspondence with the Wigner function)  the rate will be UV-finite for all physically normalized states. These are  finite energy inertial states in which  $f(k)$  falls fast enough
at large $k$. So, for the transition rate to turn negative, we require the correction term to be negative and dominant over the vacuum thermal part. It will be interesting to investigate some relevant physical cases in which this happens, which we aim to report in a subsequent publication.

\section{Conclusions}

The key non-trivial result which we obtain on combining the principles of general covariance and quantum theory is that the  particle content of a quantum state is not generally covariant. While this result, a priori, has nothing to do with gravity, it really acquires significance only in a curved spacetime. This is because, in flat spacetime, there is always a preferred foliation associated with inertial observers which can be used to define the particle content. Since such a global inertial frame does not exist in an arbitrary curved spacetime, one faces all sorts of difficulties in using concepts wedded to the particle content of a quantum state in a general curved spacetime.

Given this fact, it certainly makes sense to study the corresponding situation in non-inertial frames in flat spacetime itself. The classic example of this is the study of Rindler frame with a preferred foliation adapted for uniformly accelerated observer. Two key results which arise from such a study are the following:

(a) The quantum field theory based on the Rindler foliation of the flat spacetime interprets the inertial vacuum state as a thermal state. 

(b) An Unruh-DeWitt detector moving along a uniformly accelerated trajectory will detect a thermal spectrum of particles in the inertial vacuum. It is also well known \cite{DetvsBGLB} that the correspondence between the results (a) and (b) is a special feature of Rindler foliation and does not extend to more general foliations.

While there is extensive literature on the particle content of inertial vacuum state, detailed discussion of the particle content of inertial non-vacuum states when viewed from the Rindler frame is sparse. The only literature in similar spirit, the authors could find  were \cite{AudretschMuller}, \cite{PaddyTP}. In this paper, we analyze this situation, hopefully in a comprehensive manner, both from the point of view of quantum field theory and from that of detector response. Some of the earlier results can be obtained as special limits of the case discussed here. Our key new results can be summarized as follows.

\begin{itemize}

\item We found it convenient to encode the information about the inertial quantum state in terms of an effective field $\Phi_{\rm eff}$ which depends on a particular choice of foliation \textit{and} the quantum state. All physical quantities --- including the two-point correlation functions of the field, the particle content of the quantum state as perceived in any other foliation and the response of the detector in any arbitrary trajectory --- can be expressed in terms of $\Phi_{\rm eff}$ in rather simple and elegant manner.

We believe that this formalism will be useful in analyzing a host of other phenomena involving quantum field theory in curved spacetime. By using the same formalism with ``in'' and ``out'' vacua or even with vacuum states defined at two instance of time, one can obtain similar and useful results in contexts like particle production in expanding universe, Schwinger effect and black hole evaporation. These avenues will be explored in a future publication.

\item Having developed the formalism, we applied it to study the particle content of inertial \textit{non-vacuum } states (and in particular, superposition of one-particle states) as viewed in the Rindler foliation. For any state, the particle content has  a thermal part corresponding to the inertial vacuum state plus additional contributions. Since the former is well known in literature, we concentrated on the additional contributions throughout the paper. We found that, when the inertial quantum state is normalizable, the corrections to the thermal nature die down at high frequencies as measured in the Rindler frame. In other words, the high frequency sector of the Rindler frame perceives dominantly a thermal spectrum  in any normalizable inertial quantum state. (This result has important implications in other contexts like black hole evaporation which we hope to return to in a future work.) 

\item When the inertial quantum state is not normalizable, the situation changes drastically. For example, if the inertial state corresponds to a single particle excitation with a definite momentum $\textbf{k}$ such a state is not normalizable according to our definition. When perceived from the Rindler frame, the particle content of such a state has a \textit{constant} contribution at all high frequencies. It is as though a single particle with a definite frequency in the inertial frame is uniformly spread over all high frequencies in the Rindler frame. This result is curious and surprising and deserves further study. 

\item As we mentioned earlier, the particle content of a quantum state, determined through quantum field theory adapted to a particular foliation, may not always agree with the one determined by detector response. But since these two are known to coincide for the \textit{vacuum} state viewed from a Rindler frame, we investigated the situation as regards excited states in the inertial frame. We find that the result holds even in this context when the rate of transition of the detector is defined in a sensible manner. We stress that the validity of this result for excited states is not obvious and hence this analysis is somewhat non-trivial.

\item We also clarified  the notion of rate of transition of the detector when the Whitman function does not possess time translation invariance. There are two natural ways of defining the rate of transition in such a context which have been used to in the literature. We showed  by straightforward algebra that the result of these two procedures, in general, must be different. In view of this, we analyzed the rate of transition of a detector in a general inertial state using both the approaches. We found that there is a clear correspondence between the asymptotic (i.e., high frequency) properties of the detector transition rate and that obtained from quantum field theory in Rindler foliation.  

\end{itemize}

As indicated earlier, there are several further avenues of exploration which are suggested by our analysis. In particular, the question of black hole formation when the initial quantum state is not a vacuum state is of particular interest. Our preliminary study shows that the results obtained in this paper can be directly generalized to this case thereby allowing us to compute the correction to the thermal spectrum in black hole evaporation. This has important implications for the black hole information paradox and will be addressed in a future publication.

\section*{Acknowledgements}

The research of TP is partially supported by the J.C. Bose research grant of the Department of Science and Technology,
Government of India. KL wishes to thank Satyabrata Sahu for useful discussions. We thank the referee for several useful comments.

\vspace{2 in}

\begin{appendices}

\section{: Effective fields for a general state} 

A general state will be a superposed one for various $n$-tuply excited states. A general $n$-tuple excited state will be given as
\bea
|\Psi \rangle = \frac{1}{\sqrt{n!}}\sum_{k_1,..,k_n}f^1(k_1)...f^n(k_n) \hat{a}^{\dagger}_{k_1}..\hat{a}^{\dagger}_{k_n}|0\rangle_u
\eea
where $f^1,..f^n$ are basis functions for a $n$-variable function. For the field in $u$ modes
\bea
\phi = \sum_k (\hat{a}_k u_k + \hat{a}^{\dagger}_{k} u_k^*),
\eea
the two point function correction term will be
\bea
{\cal C}(x,y)= \frac{1}{n!}\sum_{k_1,..,k_n}\sum_{k_1',..,k_n'}{}_u\langle 0|\hat{a}_{k_1}..\hat{a}_{k_n}\sum_{kk'}(\hat{a}_k u_k(x) + \hat{a}^{\dagger}_{k} u_k^*(x))(\hat{a}_k' u_{k'}(y) + \hat{a}^{\dagger}_{k'} u_{k'}^*(y))\hat{a}^{\dagger}_{k_1}..\hat{a}^{\dagger}_{k_n}|0\rangle_u \times \nonumber\\
f^1(k_1)...f^n(k_n)f^{1*}(k'_1)...f^{n*}(k'_n)
-{}_u\langle 0| \sum_k (\hat{a}_k u_k(x) + \hat{a}^{\dagger}_{k} u_k^*(x))\sum_k' (\hat{a}_k' u_{k'}(y) + \hat{a}^{\dagger}_{k'} u_{k'}^*(y))|0\rangle_u. \label{CorrectiveCorrelationA}
\eea
The typical structure of non vanishing terms will be of the form 
\bea
\sim \frac{1}{n!}\sum_{k_1,..,k_n}\sum_{k_1',..,k_n'} \delta(k'-k_1)\delta(k'-k_2)\delta(k_3-k_4')...\delta(k_n-k_1')f^1(k_1)...f^n(k_n)f^{1*}(k'_1)...f^{n*}(k'_n),
\eea
which will have a contribution in \ref{CorrectiveCorrelationA} of the type
\bea
\sim \frac{1}{n!}\sum_{k_2,..,k_n}\sum_{k_1',k_3',.,k_n'} f^2(k_2)...f^n(k_n)f^{1*}(k'_1)f^{3*}(k'_3)..f^{n*}(k'_n) \delta(k_2-k_3')...\delta(k_n-k_1')\times \nonumber\\
\sum_{k_1,k_2'}(f^1(k_1)u_{k_1}(x))(f^{2*}(k_2')u_{k_2'}^*(y)).
\eea
Using the finiteness of the inner products of the states, we can write this term as proportional to $\Phi_1(x)\Phi_2^*(y)$, 
where $\Phi_i=\sum_k f^i_ku_k. $
Therefore, the effective correlation function is a sum of products of various $\Phi$s. And again, all the relevant quantities can be evaluated using them. The full information about $n$ different $f^i(k)$ is captured by as many $\Phi_i$s. A most general state will be a superposition of different $n$-tuple states  $|n\rangle$ and the effective correlation function will have terms squeezed between different $|n\rangle$. However, one can verify that the effective correlation function gets terms of previous kind only, with addition of some new $\Phi$ fields. A demonstration for Rindler case is presented in Appendices B and C.

\section{: Particle content for n-tuple state}\label{numberexptappendix}
The most general expression of the correction part is given as
\bea
N_{\Omega} =
\int_0^{\infty} d \tilde{\omega}\int_0^{\infty} d \tilde{\omega}'
[
\alpha_{\Omega \tilde{\omega}}\alpha^*_{\Omega \tilde{\omega}'}\langle\Psi|\hat{a}^{\dagger}(\tilde{\omega})\hat{a}(\tilde{\omega}')|\Psi\rangle+
\beta_{\Omega \tilde{\omega}}\beta^*_{\Omega \tilde{\omega}'}\langle\Psi|\hat{a}^{\dagger}(\tilde{\omega}')\hat{a}(\tilde{\omega})|\Psi\rangle \nonumber\\
-
\alpha_{\Omega \tilde{\omega}}\beta^*_{\Omega \tilde{\omega}'}\langle\Psi|\hat{a}^{\dagger}(\tilde{\omega})\hat{a}^{\dagger}(\tilde{\omega}')|\Psi\rangle-
\beta_{\Omega \tilde{\omega}}\alpha^*_{\Omega \tilde{\omega}'}\langle\Psi|\hat{a}(\tilde{\omega})\hat{a}(\tilde{\omega}')|\Psi\rangle
].\label{NEx3}
\eea

A typical $n$-tuply excited state in Minkowski can be given as

\bea
|\Psi\rangle= \int_0^{\infty}\prod_{i=1}^n \frac{d \omega_i}{\sqrt{2\pi\omega_i}} F(\omega_1,...\omega_n) \hat{a}^{\dagger}(\omega_i)|0 \rangle_M, \label{nPS}
\eea
with
\bea
F(\omega_1,...\omega_n)=\sum_{i_1,i_2..,i_n}c^{i_1,i_2..,i_n}f_{i_1}(\omega_1)...f_{i_n}(\omega_n), \label{FDecomp}
\eea
where $f_{i_1}(\omega) $ can be considered as basis functions in the function space.\\
for $n-$ tuple state, we are are left with finding the expression for first two terms in \ref{NEx3}, which ultimately boils down on finding $\langle\Psi|\hat{a}^{\dagger}(\tilde{\omega})\hat{a}(\tilde{\omega}')|\Psi\rangle $ for the state given by \ref{nPS}. Thus we have
\bea
\langle\Psi|\hat{a}^{\dagger}(\tilde{\omega})\hat{a}(\tilde{\omega}')|\Psi\rangle=
\frac{1}{n!}\left(\prod_{i,j}\int \int\frac{ d\omega_i}{\sqrt{2\pi\omega_i}} \frac{ d\omega_j'}{\sqrt{2\pi\omega_j'}} \right)F(\omega_1,...\omega_n)F^*(\omega_1',...\omega_n')\times \nonumber\\
{}_M\langle 0|[\prod_m\hat{a}(\omega_m')]\hat{a}^{\dagger}(\tilde{\omega})\hat{a}(\tilde{\omega}')[\prod_l\hat{a}^{\dagger}(\omega_l)] |0 \rangle_M. \label{NptEx}
\eea
Now, using the commutation rules we can write
\bea
\hat{a}(\tilde{\omega}')\hat{a}^{\dagger}(\omega_1)..\hat{a}^{\dagger}(\omega_n)=\delta(\tilde{\omega}'-\omega_1)\hat{a}^{\dagger}(\omega_2)..\hat{a}^{\dagger}(\omega_n)+\delta(\tilde{\omega}'-\omega_2)\hat{a}^{\dagger}(\omega_1)\hat{a}^{\dagger}(\omega_3)..\hat{a}^{\dagger}(\omega_n)+
.....+\nonumber\\
\delta(\tilde{\omega}'-\omega_n)\hat{a}^{\dagger}(\omega_2)..\hat{a}^{\dagger}(\omega_{n-1})+\hat{a}^{\dagger}(\omega_1)..\hat{a}^{\dagger}(\omega_n)\hat{a}(\tilde{\omega}') 
\eea
Therefore,
\bea
{}_M\langle 0|\hat{a}(\omega_n')..\hat{a}(\omega_1')\hat{a}^{\dagger}(\tilde{\omega})\hat{a}(\tilde{\omega}')\hat{a}^{\dagger}(\omega_1)..\hat{a}^{\dagger}(\omega_n) |0 \rangle_M =\delta(\tilde{\omega}'-\omega_1)\delta(\tilde{\omega}-\omega_1'){}_M\langle 0|\hat{a}(\omega_n')..\hat{a}(\omega_2')\hat{a}^{\dagger}(\omega_2)..\hat{a}^{\dagger}(\omega_n) |0 \rangle_M + \nonumber\\ \delta(\tilde{\omega}'-\omega_1)\delta(\tilde{\omega}-\omega_2'){}_M\langle 0|\hat{a}(\omega_n')..\hat{a}(\omega_3')\hat{a}(\omega_1')\hat{a}^{\dagger}(\omega_2)..\hat{a}^{\dagger}(\omega_n) |0 \rangle_M +... ...\nonumber\\
+\delta(\tilde{\omega}'-\omega_2)\delta(\tilde{\omega}-\omega_1'){}_M\langle 0|\hat{a}(\omega_n')..\hat{a}(\omega_2')\hat{a}^{\dagger}(\omega_1)\hat{a}^{\dagger}(\omega_3)..\hat{a}^{\dagger}(\omega_n) |0 \rangle_M + \nonumber\\
.... .... \label{Preexpectation}
\eea
In the above expression, in the right hand side the terms squeezed between $|0 \rangle_M $ give rise to various delta functions involving various $\omega$ and $\omega'$. Therefore, in \ref{NptEx} we use \ref{FDecomp} and realize that ultimately we obtain a term proportional to
\bea
\sum_{ij}\tilde{c}^{i j}\frac{1}{\sqrt{2\tilde{\omega}2\tilde{\omega}'}}f_i(\tilde{\omega})f_j^*(\tilde{\omega}'), \label{Expectation}
\eea
for some functions $f_i$ such that $f_i(\omega)/\sqrt{2\omega}$ is square integrable over $\omega \in (0,\infty]$. All other $f_j$s in $F$ as shown in \ref{FDecomp}, get involved in inner products among themselves to give rise to some finite numbers. All this can be easily seen by an example of a doubly excited state.\\
Let us have a state given by
\bea
F(\omega_1,\omega_2)=\frac{1}{\sqrt{2\tilde{\omega}2\tilde{\omega}'}}(f_1(\omega_1)f_2(\omega_2)+\tilde{f}_1(\omega_1)\tilde{f}_2(\omega_2)),
\eea
for simple demonstration, with all $f_i(\omega)/\sqrt{2\omega},\tilde{f}_i(\omega)/\sqrt{2\omega}$ being square integrable over $\omega \in (0,\infty]$. \\

In this case, 
\bea 
{}_M\langle 0|\hat{a}(\omega_2')\hat{a}(\omega_1')\hat{a}^{\dagger}(\tilde{\omega})\hat{a}(\tilde{\omega}')\hat{a}^{\dagger}(\omega_1)\hat{a}^{\dagger}(\omega_2) |0 \rangle_M =\delta(\tilde{\omega}'-\omega_1)\delta(\tilde{\omega}-\omega_1'){}_M\langle 0|\hat{a}(\omega_2')\hat{a}^{\dagger}(\omega_2)|0 \rangle_M + \nonumber\\
\delta(\tilde{\omega}'-\omega_2)\delta(\tilde{\omega}-\omega_1'){}_M\langle 0|\hat{a}(\omega_2')\hat{a}^{\dagger}(\omega_1 )|0 \rangle_M +
\delta(\tilde{\omega}'-\omega_1)\delta(\tilde{\omega}-\omega_2'){}_M\langle 0|\hat{a}(\omega_1')\hat{a}^{\dagger}(\omega_2 )|0 \rangle_M +\nonumber\\
\delta(\tilde{\omega}'-\omega_2)\delta(\tilde{\omega}-\omega_2'){}_M\langle 0|\hat{a}(\omega_1')\hat{a}^{\dagger}(\omega_1 )|0 \rangle_M,\label{Example5.1}
\eea
which is the analog of the \ref{Preexpectation} which was defined for the general case, which again using the commutation relations become simple as.
\bea
\langle\Psi|\hat{a}^{\dagger}(\tilde{\omega})\hat{a}(\tilde{\omega}')|\Psi\rangle= \int \frac{d\omega_1d\omega_2d\omega_1'd\omega_2'}{\sqrt{2\omega_12\omega_22\omega_1'2\omega_2'}}F(\omega_1,\omega_2)F^*(\omega_1',\omega_2')\times 
[\delta(\tilde{\omega}'-\omega_1)\delta(\tilde{\omega}-\omega_1')\delta(\omega_2-\omega_2') \nonumber\\
+\delta(\tilde{\omega}'-\omega_2)\delta(\tilde{\omega}-\omega_1')\delta(\omega_1-\omega_2')+ \delta(\tilde{\omega}'-\omega_1)\delta(\tilde{\omega}-\omega_2')\delta(\omega_2-\omega_1')+\delta(\tilde{\omega}'-\omega_2)\delta(\tilde{\omega}-\omega_2')\delta(\omega_1-\omega_1')].
\eea
Therefore, we ultimately have
\bea
\frac{1}{\sqrt{2\tilde{\omega}2\tilde{\omega}'}}\int\frac{d\omega}{2\omega}\left[F(\tilde{\omega}',\omega)F^*(\tilde{\omega},\omega) +
F(\omega,\tilde{\omega}')F^*(\tilde{\omega},\omega)+F(\tilde{\omega}',\omega)F^*(\omega,\tilde{\omega})+F(\omega,\tilde{\omega}')F^*(\omega,\tilde{\omega})
\right]. \label{Example5.2}
\eea
For further demonstration, we take one of the term (say second) in the parenthesis of the integrand. Calculations for all other terms will be similar.
\bea
\int\frac{d\omega}{2\omega}F(\omega,\tilde{\omega}')F^*(\tilde{\omega},\omega)=\frac{1}{2}\int\frac{d\omega}{2\omega}[f_1(\omega)f_2(\tilde{\omega}')+\tilde{f}_1(\omega)\tilde{f}_2(\tilde{\omega}') ][f_1^*(\tilde{\omega})f_2^*(\omega)+\tilde{f}_1^*(\tilde{\omega})\tilde{f}_2^*(\omega) ], \nonumber\\
=\frac{1}{2}\left[f_2(\tilde{\omega}')f_1^*(\tilde{\omega})\int\frac{d\omega}{2\omega}f_1(\omega)f_2^*(\omega)+ f_2(\tilde{\omega}')\tilde{f}_1^*(\tilde{\omega})\int\frac{d\omega}{2\omega}f_1(\omega)\tilde{f}_2^*(\omega)\right]\nonumber\\
+\frac{1}{2}\left[
\tilde{f}_2(\tilde{\omega}')f_1^*(\tilde{\omega})\int\frac{d\omega}{2\omega}\tilde{f}_1(\omega)f_2^*(\omega)+ \tilde{f}_2(\tilde{\omega}')\tilde{f}_1^*(\tilde{\omega})\int\frac{d\omega}{2\omega}\tilde{f}_1(\omega)\tilde{f}_2^*(\omega)\right] \label{Example5Final}.
\eea
If we again do the transformation $\log{\frac{\omega}{a}}=t$, we can easily see that every integral in \ref{Example5Final} becomes an inner product between different square integrable functions and is finite. Similar contributions will come from all the terms in \ref{Example5.2}. Therefore, the expression in \ref{Example5.1} turns into one like \ref{Expectation}.

Now when we use the integrations of terms in \ref{NptEx} with $\alpha^*_{\Omega \tilde{\omega}} \alpha_{\Omega \tilde{\omega}'}$ (or $\beta^*_{\Omega \tilde{\omega}} \beta_{\Omega \tilde{\omega}'}$) for obtaining contributions in \ref{NEx3}, we again obtain that the correction term being a sum of (products of) Fourier transforms of various square integrable functions, with respect to $\Omega/a$, as previously seen in \ref{FourierTransform1} and \ref{FourierTransform2A}. Evidently the correction terms decay down for large $\Omega/a$, at least as fast as $(\Omega/a)^{-1+\epsilon}$. Therefore, each term in \ref{Expectation} gives rise to product of Fourier transforms and is decaying. Thus, decaying nature of correction terms is true for arbitrary $n-$ tuple states too. Now, we can generalize this for most general arbitrary states (superpositions of various particle states). Evidently, the important terms in that case will be the ones generating correlations, i.e. the last two terms in \ref{NEx3}, which we deal with next.

\section{: Correlation terms in a general superposition state}

In a general superposed states, there will be terms different from those considered above, which are correlation terms between an $n-$ tuple and $(n-2)$ tuple state, i.e.
\bea
\langle n|\hat{a}^{\dagger}(\tilde{\omega})\hat{a}(\tilde{\omega}')|n-2\rangle,
\eea
where $|n\rangle$  is a general $n-$ tuple state considered in \ref{nPS}. For demonstration purpose we will consider a particular term arising out of various components in superposition. Its generic form will be
\bea
\langle n|\hat{a}^{\dagger}(\tilde{\omega})\hat{a}^{\dagger}(\tilde{\omega}')|n-2\rangle \propto 
\left(\prod_{i,j}\int \int\frac{ d\omega_i}{\sqrt{2\pi\omega_i}} \frac{ d\omega_j'}{\sqrt{2\pi\omega_j'}} \right)f_1^*(\omega_1)...f_n^*(\omega_n)
\tilde{f}_1(\omega_1')...\tilde{f}_{n-2}(\omega_{n-2}') \times \nonumber\\
{}_M\langle 0|\hat{a}(\omega_n)..\hat{a}(\omega_1)\hat{a}^{\dagger}(\tilde{\omega})\hat{a}^{\dagger}(\tilde{\omega}')\hat{a}^{\dagger}(\omega_1')..\hat{a}^{\dagger}(\omega_{n-2}') |0 \rangle_M. 
\eea

Again, shifting, $\hat{a}^{\dagger}(\tilde{\omega})$ and $\hat{a}^{\dagger}(\tilde{\omega}')$ to the left  to right will give various delta functions involving 
$\tilde{\omega}$ and $\tilde{\omega}'$  with various $\omega_i$s. Thus again we are left with terms proportional to 
\bea
\sum_{ij}\tilde{c}^{i j}\frac{1}{\sqrt{2\tilde{\omega}2\tilde{\omega}'}}f_i^*(\tilde{\omega})f_j^*(\tilde{\omega}'). \label{Expectation2}
\eea
Therefore, yet again in \ref{NEx3} we have Fourier transforms of two square integrable functions which is decaying in $\Omega/a$. Additionally since such terms are integrated with $\alpha^*_{\Omega \tilde{\omega}}\beta_{\Omega \tilde{\omega}'}$, the Fourier transform is multiplied with
\bea
\frac{\Omega}{4\pi^2a^2}\Gamma\left[\frac{i\Omega}{a}\right] \Gamma\left[-\frac{i\Omega}{a}\right]=\frac{1}{4\pi a}\cosech{\frac{\pi\Omega}{a}}.
\eea

Therefore, such terms will decay even faster than the previously considered terms. Again, for clarity we will quickly go through an example with a state
\bea
|\Psi\rangle = c_0|0\rangle + c_1\int \frac
{d\omega}{\sqrt{2\omega}} f(\omega)\hat{a}^{\dagger}(\omega)|0\rangle + c_2 \int \int \frac{d\omega}{\sqrt{2\omega}}\frac{d\omega'}{2\sqrt{\omega'}}f_1(\omega)f_2(\omega')\hat{a}^{\dagger}(\omega)\hat{a}^{\dagger}(\omega')|0\rangle.
\eea
Since we already know the profiles of first two terms in \ref{NEx3}, we will consider the correlation terms which are of type $\langle\Psi|\hat{a}^{\dagger}(\tilde{\omega})\hat{a}^{\dagger}(\tilde{\omega}')|\Psi\rangle$ and $\langle\Psi|\hat{a}(\tilde{\omega})\hat{a}(\tilde{\omega}')|\Psi\rangle$. Therefore, such a non-vanishing term will be
\bea 
\langle\Psi|\hat{a}^{\dagger}(\tilde{\omega})\hat{a}^{\dagger}(\tilde{\omega}')|\Psi\rangle &=& c_0c_2^*\int \int \frac{d\omega_1}{\sqrt{2\omega_1}}\frac{d\omega_2}{\sqrt{2\omega_2}}f_1^*(\omega_1)f_2^*(\omega_2)\langle 0|a(\omega_1)a(\omega_2)\hat{a}^{\dagger}(\tilde{\omega})\hat{a}^{\dagger}(\tilde{\omega}')|0\rangle, \\
&=& c_0c_2^*\int \int \frac{d\omega_1}{\sqrt{2\omega_1}}\frac{d\omega_2}{\sqrt{2\omega_2}}f_1^*(\omega_1)f_2^*(\omega_2)[\delta(\omega_1-\tilde{\omega})\delta(\omega_2-\tilde{\omega}')+\delta(\omega_1-\tilde{\omega}')\delta(\omega_2-\tilde{\omega})], \\
&=& c_0c_2^*\frac{1}{\sqrt{4\tilde{\omega}\tilde{\omega}'}}[f_1^*(\tilde{\omega})f_2^*(\tilde{\omega}')+f_1^*(\tilde{\omega}')f_2^*(\tilde{\omega})].
\eea

In \ref{NEx3}, these terms get integrated over $\tilde{\omega}$ and $\tilde{\omega}'$ with $\alpha^*_{\Omega \tilde{\omega}}\beta_{\Omega \tilde{\omega}'}$ and as we discussed before  will additionally contribute a factor $\frac{1}{4\pi a}\cosech{\frac{\pi\Omega}{a}}$,
apart from the Fourier transforms of $\frac{f_1^*(\omega)}{\sqrt{2\omega}}$ and  $\frac{f_2^*(\omega)}{\sqrt{2\omega}}$, which are already decaying functions for large $\Omega/a$. A very similar analysis shows similar result for $\langle\Psi|\hat{a}(\tilde{\omega})\hat{a}(\tilde{\omega}')|\Psi\rangle$,
which in fact will be the complex conjugate of this term. Thus, we see that the correlation terms decay even faster for large $\Omega/a$. This can also be generalized for other higher excitation states in a similar fashion.
\\

Now, we have generalized our result for all physically meaningful, normalizable states. For such states, there will always be a length scale below which the Rindler observer will not detect any significant departure from the black-body profile. However, those observers having access to larger and larger modes, will observe progressively larger corrections.

\section{: Detector response versus Number expectation}

We saw that for the one particle state, the number expectation correction for the left-moving modes corresponds to the detector response upto a factor $\pi$. It is important to note that this factor is always present between the number expectation value and the detector response, if the distribution is only among one set of modes (say left-moving modes only). In vacuum response, there is in fact a $2 \pi$ factor discrepancy, since in that case both the left and right moving modes contribute to the detector response. This factor is adjusted by the normalization of the $\omega$ integral. Similarly, if the number expectation is calculated for both the left and right moving modes (see Appendix E) with the above-mentioned normalization, it will again match with the detector response. Thus the correspondence between (a) mean number of particles defined using the Bogoliubov coefficients appropriate for the foliation and (b) transition of detectors continue to hold even for the one particle (and also for general excited) states. We quickly demonstrate these points for the vacuum response. 

The  first order transition probability for the vacuum state will be given as
\bea 
P_{\text{vac}} = \int_{-\infty}^{\infty} \int_{-\infty}^{\infty} d \tau d \tau' e^{-i \Delta E (\tau-\tau')} \int_{-\infty}^{\infty} d k
u_k (\tau) u_k^*(\tau'),
\eea
with 
\bea
u_k(\tau) = \frac{1}{\sqrt{4 \pi |k|}}e^{-i|k|t(\tau)+i k x(\tau)}, \label{MinkowskimodeinRindler}
\eea
for the Rindler trajectory \ref{RindlerTraj}. We first evaluate the $\tau$ integrals 
\bea
\int_{-\infty}^{\infty} \int_{-\infty}^{\infty} d \tau d \tau' e^{-i \Delta E (\tau-\tau')} \int_{-\infty}^{\infty} d k
u_k (\tau) u_k^*(\tau')= \int_{-\infty}^{\infty} \int_{-\infty}^{\infty} d \tau d \tau' \int_{0}^{\infty} d k\left[ u_k (\tau) u_k^*(\tau') +  u_{-k} (\tau) u_{-k}^*(\tau') \right].
\eea

Using \ref{MinkowskimodeinRindler} and \ref{tauintegral1} we  get

\bea
P_{\text{vac}} = \int_{0}^{\infty}\frac{dk}{k}\frac{1}{2\Delta E a}\left[\frac{1}{e^{\frac{2 \pi \Delta E}{a}}-1} + \frac{1}{e^{\frac{2 \pi \Delta E}{a}}-1} \right],
\eea
where each term is a contribution of left and right moving modes respectively, while
\bea
\int_{0}^{\infty}\frac{d \omega }{\omega} |\beta_{\Omega \omega}|^2 = \int_{0}^{\infty}\frac{d \omega }{2 \pi \omega a} \frac{1}{e^{\frac{2 \pi \Omega}{a}}-1}.
\eea
Thus we see that if we concentrate only on the left moving mode in Rindler it differs from the corresponding one in the detector response by a factor $1/\pi$. However, the full contribution coming from the left as well as right moving modes makes the difference factor $1/2\pi$ which is compensated by a $2\pi \delta(0)$ contribution from the $\omega$ integral. Similarly, in Appendix E, we will see that the if the distribution for a general state is specified to be only among the left-moving modes, the corresponding terms in the number expectation value and the detector response differ by $1/\pi$. This discrepancy is basically generated by the relation
\bea
\int_{-\infty}^{\infty} \int_{-\infty}^{\infty} d \tau d \tau' e^{-i \Omega (\tau-\tau')}u_k (\tau) u_k^*(\tau') = \frac{\pi}{\Omega} |\beta_{\Omega \omega_k}|^2.
\eea

\section{: Detector response for Rindler trajectories}

The two point function in the singly excited state is given as

\bea
\langle 1_M | \phi(x)\phi(y) | 1_M \rangle = \int \frac{dk}{2\pi}e^{-ik\cdot(x-y)}+ \int \frac{dk}{2\pi}\int \frac{dk'}{2\pi}\frac{f(k)}{2\omega_k}\frac{f^*(k')}{2\omega_k'}e^{-ik\cdot x+ik'\cdot y} +\text{c.c.} \label{TwoPointFunction}
\eea

We can easily identify the first term as the standard vacuum contribution, which is a standard feature of the two point function. So for the detector too, there always be a vacuum contribution which is the thermal part. Henceforth we will concentrate on the correction terms only.

We first analyze the momentum integrals in the correction term

\bea
\int \frac{dk}{2\pi}\int \frac{dk'}{2\pi}\frac{f(k)}{2\omega_k}\frac{f^*(k')}{2\omega_k'}e^{-ik\cdot x+ik'\cdot y}+\text{c.c.}=
\int \frac{dk}{2\pi}\int \frac{dk'}{2\pi}\frac{f(k)}{2\omega_k}\frac{f^*(k')}{2\omega_k'}e^{-i(\omega_k t-kx) + i (\omega_k't'-k'x')}+\text{c.c.} \label{b4dotproduct}
\eea

For massless scalar field $\omega_k=|k|$. Thus we can expand  \ref{b4dotproduct} as
\bea
\int_0^{\infty}\frac{dk}{4\pi k}\int_0^{\infty}\frac{dk'}{4\pi k'}f(k)f^*(k')e^{-i[k(t-x) - i k'(t'-x')]}+\int_0^{\infty}\frac{dk}{4\pi k}\int_0^{\infty}\frac{dk'}{4\pi k'}f^*(k)f(k')e^{i[k(t-x) - k'(t'-x')]}\nonumber\\
+
\int_0^{\infty}\frac{dk}{4\pi k}\int_0^{\infty}\frac{dk'}{4\pi k'}f(k)f^*(-k')e^{-i[k(t-x) - i k'(t+x')]}+\int_0^{\infty}\frac{dk}{4\pi k}\int_0^{\infty}\frac{dk'}{4\pi k'}f^*(k)f(-k')e^{-i[k(t-x) - k'(t'+x')]}\nonumber\\
+
\int_0^{\infty}\frac{dk}{4\pi k}\int_0^{\infty}\frac{dk'}{4\pi k'}f(-k)f^*(k')e^{-i[k(t+x) - i k'(t'-x')]}+\int_0^{\infty}\frac{dk}{4\pi k}\int_0^{\infty}\frac{dk'}{4\pi k'}f^*(-k)f(k')e^{i[k(t+x) - k'(t'-x')]}\nonumber\\
+
\int_0^{\infty}\frac{dk}{4\pi k}\int_0^{\infty}\frac{dk'}{4\pi k'}f(-k)f^*(-k')e^{-i[k(t+x) - i k'(t+x')]}+\int_0^{\infty}\frac{dk}{4\pi k}\int_0^{\infty}\frac{dk'}{4\pi k'}f^*(-k)f(-k')e^{i[k(t+x) - k'(t'+x')]}. \label{CorwithDot}
\eea

Now we can evaluate the full correction term for the Rindler trajectory \ref{RindlerTraj} and write the expression in $T,s$ co-ordinates  (defined previously) as,
\bea
P =\int ds e^{-i \Delta E s}[I_1+I_2+I_3+I_4+ c.c.],
\eea
with
\bea
\int ds e^{-i \Delta E s} I_1\equiv \int \int dT ds e^{-i \Delta E s}\int_0^{\infty}\frac{dk}{4\pi k}\int_0^{\infty}\frac{dk'}{4\pi k'}f(k)f^*(k')e^{-i[k(t-x) - i k'(t'-x')]}= \nonumber\\
\int \int dT ds e^{-i \Delta E s} \int_0^{\infty} \frac{dk}{2\pi}\int_0^{\infty} \frac{dk'}{2\pi}\frac{f(k)}{2k}\frac{f^*(k')}{2k'}e^{\left(+i\frac{k}{a}e^{-a(2T+s)/2}-i\frac{k}{a}e^{-a(2T-s)/2}\right)}. \label{correction1}
\eea

We first evaluate the $s$ integral, again with making a transformation $e^{-a s/2} = z$

\bea
\int_{-\infty}^{\infty} dse^{-i \Delta E s}  e^{+i\frac{k}{a}e^{-a(2T+s)/2}-i\frac{k}{a}e^{-a(2T-s)/2}} =\frac{2}{a}\int_0^{\infty}dz z^{\frac{2i\Delta E}{a}-1}
e^{-i\frac{k'}{a}e^{-a T} \frac{1}{z} +i\frac{k}{a}e^{-a T}z} \nonumber\\
=\frac{4}{a}e^{-\frac{\pi}{a} \Delta E} \left(\frac{k}{a}e^{-a T}\right)^{-i\frac{\Delta E}{a}}\left(\frac{k'}{a}e^{-a T}\right)^{i\frac{\Delta E}{a}}
K_{-2i\frac{\Delta E}{a}}\left[2\frac{\sqrt{k k'}}{a}e^{-a T}\right]. \label{integration1}
\eea

Therefore, substituting \ref{integration1} in \ref{correction1}, we get

\bea
\int ds e^{-i \Delta E s} I_1&=&\int dT \int_0^{\infty} \frac{dk}{2\pi}\int_0^{\infty} \frac{dk'}{2\pi}\frac{f(k)}{2k}\frac{f^*(k')}{2k'}\left[\frac{4}{a}e^{-\frac{\pi}{a}\Delta E} \left(\frac{k}{a}e^{-a T}\right)^{-i\frac{\Delta E}{a}}\left(\frac{k'}{a}e^{-a T}\right)^{i\frac{\Delta E}{a}}
K_{-2i\frac{\Delta E}{a}}\left[2\frac{\sqrt{k k'}}{a}e^{-a T}\right]\right] \label{integration2} \nonumber\\
\int ds e^{-i \Delta E s} I^*_1&=&\int dT \int_0^{\infty} \frac{dk}{2\pi}\int_0^{\infty} \frac{dk'}{2\pi}\frac{f^*(k)}{2k}\frac{f(k')}{2k'}\left[\frac{4}{a}e^{\frac{\pi}{a}\Delta E} \left(\frac{k}{a}e^{-a T}\right)^{-i\frac{\Delta E}{a}}\left(\frac{k'}{a}e^{-a T}\right)^{i\frac{\Delta E}{a}}
K_{-2i\frac{\Delta E}{a}}\left[2\frac{\sqrt{k k'}}{a}e^{-a T}\right]\right]. \label{integration2A}\nonumber\\
\eea

Similarly, we obtain

\bea
\int ds e^{-i \Delta E s} I_2 &\equiv&  \int \int dT ds e^{-i \Delta E s} \int_0^{\infty}\frac{dk}{4\pi k}\int_0^{\infty}\frac{dk'}{4\pi k'}f(k)f^*(-k')e^{-i[k(t-x) - i k'(t+x')]} \nonumber\\
&=&\int dT \int_0^{\infty} \frac{dk}{2\pi}\int_0^{\infty} \frac{dk'}{2\pi}\frac{f(k)}{2k}\frac{f^*(-k')}{2k'}\left[\frac{2}{a}e^{-\frac{\pi}{a}\Delta E}\left(\frac{k}{a}e^{-aT}+ \frac{k'}{a}e^{aT}\right)^{-2i\frac{\Delta E}{a}}\Gamma\left[2 i\frac{\Delta E}{a} \right] \right]. \label{integration3}\\
\int ds e^{-i \Delta E s} I^*_2&=&\int dT \int_0^{\infty} \frac{dk}{2\pi}\int_0^{\infty} \frac{dk'}{2\pi}\frac{f^*(k)}{2k}\frac{f(-k')}{2k'}\left[\frac{2}{a}e^{\frac{\pi}{a}\Delta E}\left(\frac{k}{a}e^{-aT}+ \frac{k'}{a}e^{aT}\right)^{-2i\frac{\Delta E}{a}}\Gamma\left[2 i\frac{\Delta E}{a} \right] \right]. \label{integration3A}
\eea
Also,
\bea
\int ds e^{-i \Delta E s} I_3 &\equiv& \int \int dT ds e^{-i \Delta E s} \int_0^{\infty}\frac{dk}{4\pi k}\int_0^{\infty}\frac{dk'}{4\pi k'}f(-k)f^*(k')e^{-i[k(t+x) - i k'(t'-x')]} \nonumber\\
&=& \int dT \int_0^{\infty} \frac{dk}{2\pi}\int_0^{\infty} \frac{dk'}{2\pi}f(-k)f^*(k')\left[\frac{2}{a}e^{-\frac{\pi}{a}\Delta E}\left(\frac{k}{a}e^{-aT}+ \frac{k'}{a}e^{aT}\right)^{-2i\frac{\Delta E}{a}}\Gamma\left[-2 i\frac{\Delta E}{a} \right] \right],\label{integration4} \\
\int ds e^{-i \Delta E s} I^*_3 &=& \int dT \int_0^{\infty} \frac{dk}{2\pi}\int_0^{\infty} \frac{dk'}{2\pi}f^*(-k)f(k')\left[\frac{2}{a}e^{\frac{\pi}{a}\Delta E}\left(\frac{k}{a}e^{-aT}+ \frac{k'}{a}e^{aT}\right)^{-2i\frac{\Delta E}{a}}\Gamma\left[-2 i\frac{\Delta E}{a}\right] \right].\label{integration4A}
\eea
And,
\bea
\int ds e^{-i \Delta E s} I_4 &\equiv& \int \int dT ds e^{-i \Delta E s}\int_0^{\infty}\frac{dk}{4\pi k}\int_0^{\infty}\frac{dk'}{4\pi k'}f(-k)f^*(-k')e^{-i[k(t+x) - i k'(t+x')]}
\nonumber\\
&=&\int dT \int_0^{\infty} \frac{dk}{2\pi}\int_0^{\infty} \frac{dk'}{2\pi}f(-k)f^*(-k')
\left[\frac{4}{a}e^{-\frac{\pi}{a}\Delta E} \left(\frac{k}{a}e^{-a T}\right)^{i\frac{\Delta E}{a}}\left(\frac{k'}{a}e^{-a T}\right)^{-i\frac{\Delta E}{a}}
K_{-2i\frac{\Delta E}{a}}\left[2\frac{\sqrt{k k'}}{a}e^{-a T}\right]\right].\label{integration5}\nonumber\\
\int ds e^{-i \Delta E s} I^*_4 &=& \int dT \int_0^{\infty} \frac{dk}{2\pi}\int_0^{\infty} \frac{dk'}{2\pi}f^*(-k)f(-k')
\left[\frac{4}{a}e^{\frac{\pi}{a}\Delta E} \left(\frac{k}{a}e^{-a T}\right)^{i\frac{\Delta E}{a}}\left(\frac{k'}{a}e^{-a T}\right)^{-i\frac{\Delta E}{a}}
K_{-2i\frac{\Delta E}{a}}\left[2\frac{\sqrt{k k'}}{a}e^{-a T}\right]\right].\label{integration5A}\nonumber\\
\eea

For evaluating the correction in probability we also need to perform the $ T$ integrals. For that, we use following expressions

\bea
\int_{-\infty}^{\infty} dT K_{-2i\frac{\Delta E}{a}}\left[2\frac{\sqrt{k k'}}{a}e^{-a T}\right] &=& \frac{1}{4a}\Gamma\left[i\frac{\Delta E}{a} \right]\Gamma\left[- i\frac{\Delta E}{a} \right], \nonumber\\
\int_{-\infty}^{\infty} dT \left(\frac{k}{a}e^{-aT}+ \frac{k'}{a}e^{aT}\right)^{-2i\frac{\Delta E}{a}} &=& \frac{1}{2 a}\left(\frac{k}{a}\right)^{-i\frac{\Delta E}{a}}\left(\frac{k'}{a}\right)^{-i\frac{\Delta E}{a}}\frac{\Gamma\left[i\frac{\Delta E}{a} \right]^2}{\Gamma\left[2i\frac{\Delta E}{a} \right]}. \label{Simplifyingrelations}
\eea
Using, the relations \ref{Simplifyingrelations}  in \ref{integration2}-\ref{integration5A} we get the expression,
\bea
P =\frac{1}{a^2}e^{-\frac{\pi}{a}\Delta E}\Gamma\left[i\frac{\Delta E}{a} \right]\Gamma\left[- i\frac{\Delta E}{a} \right]\int_0^{\infty} \frac{dk}{2\pi}\int_0^{\infty} \frac{dk'}{2\pi}\frac{f(k)}{2k}\frac{f^*(k')}{2k'}\left(\frac{k}{a}\right)^{-i\frac{\Delta E}{a}}\left(\frac{k'}{a}\right)^{i\frac{\Delta E}{a}}+\nonumber\\
\frac{1}{a^2}e^{\frac{\pi}{a}\Delta E}\Gamma\left[i\frac{\Delta E}{a} \right]\Gamma\left[- i\frac{\Delta E}{a} \right]\int_0^{\infty} \frac{dk}{2\pi}\int_0^{\infty} \frac{dk'}{2\pi}\frac{f^*(k)}{2k}\frac{f(k')}{2k'}\left(\frac{k}{a}\right)^{-i\frac{\Delta E}{a}}\left(\frac{k'}{a}\right)^{i\frac{\Delta E}{a}}+\nonumber\\
\frac{1}{a^2}e^{-\frac{\pi}{a}\Delta E}\Gamma\left[i\frac{\Delta E}{a} \right]\Gamma\left[i\frac{\Delta E}{a} \right]\int_0^{\infty} \frac{dk}{2\pi}\int_0^{\infty} \frac{dk'}{2\pi}\frac{f(k)}{2k}\frac{f^*(-k')}{2k'}\left(\frac{k}{a}\right)^{-i\frac{\Delta E}{a}}\left(\frac{k'}{a}\right)^{-i\frac{\Delta E}{a}}+\nonumber\\
\frac{1}{a^2} e^{\frac{\pi}{a}\Delta E}\Gamma\left[i\frac{\Delta E}{a} \right]\Gamma\left[i\frac{\Delta E}{a} \right]\int_0^{\infty} \frac{dk}{2\pi}\int_0^{\infty} \frac{dk'}{2\pi}\frac{f^*(k)}{2k}\frac{f(-k')}{2k'}\left(\frac{k}{a}\right)^{-i\frac{\Delta E}{a}}\left(\frac{k'}{a}\right)^{-i\frac{\Delta E}{a}}+\nonumber\\
\frac{1}{a^2}e^{-\frac{\pi}{a}\Delta E}\Gamma\left[-i\frac{\Delta E}{a} \right]\Gamma\left[-i\frac{\Delta E}{a} \right]\int_0^{\infty} \frac{dk}{2\pi}\int_0^{\infty} \frac{dk'}{2\pi}\frac{f^*(k)}{2k}\frac{f(-k')}{2k'}\left(\frac{k}{a}\right)^{i\frac{\Delta E}{a}}\left(\frac{k'}{a}\right)^{i\frac{\Delta E}{a}}+\nonumber\\
\frac{1}{a^2} e^{\frac{\pi}{a}\Delta E}\Gamma\left[-i\frac{\Delta E}{a} \right]\Gamma\left[-i\frac{\Delta E}{a} \right]\int_0^{\infty} \frac{dk}{2\pi}\int_0^{\infty} \frac{dk'}{2\pi}\frac{f(k)}{2k}\frac{f^*(-k')}{2k'}\left(\frac{k}{a}\right)^{i\frac{\Delta E}{a}}\left(\frac{k'}{a}\right)^{i\frac{\Delta E}{a}}+\nonumber\\
\frac{1}{a^2}e^{-\frac{\pi}{a}\Delta E}\Gamma\left[i\frac{\Delta E}{a} \right]\Gamma\left[- i\frac{\Delta E}{a} \right]\int_0^{\infty} \frac{dk}{2\pi}\int_0^{\infty} \frac{dk'}{2\pi}\frac{f(-k)}{2k}\frac{f^*(-k')}{2k'}\left(\frac{k}{a}\right)^{i\frac{\Delta E}{a}}\left(\frac{k'}{a}\right)^{-i\frac{\Delta E}{a}}+\nonumber\\
\frac{1}{a^2}e^{\frac{\pi}{a}\Delta E}\Gamma\left[i\frac{\Delta E}{a} \right]\Gamma\left[- i\frac{\Delta E}{a} \right]\int_0^{\infty} \frac{dk}{2\pi}\int_0^{\infty} \frac{dk'}{2\pi}\frac{f^*(-k)}{2k}\frac{f(-k')}{2k'}\left(\frac{k}{a}\right)^{i\frac{\Delta E}{a}}\left(\frac{k'}{a}\right)^{-i\frac{\Delta E}{a}}. \label{correction41pt}
\eea
Now, if we identify 
\bea
\left(\frac{k}{a}\right)^{-i\frac{\Delta E}{a}}= e^{-i\frac{\Delta E}{a}\log{\left(\frac{k}{a}\right)}},
\eea
and again make the transformation to a new variable  $\log{\left(\frac{k}{a}\right)}=t$, all the integrals get converted into the Fourier transforms of square integrable functions in $t$, with respect to $\Delta E/a$ once again. Therefore, the correction profile, as observed by the Rindler detector has a decaying tendency for large 
$\Delta E/a$.

\end{appendices}

\end{document}